\documentclass[sigconf,nonacm, screen]{acmart}

\AtBeginDocument{%
  \providecommand\BibTeX{{%
    \normalfont B\kern-0.5em{\scshape i\kern-0.25em b}\kern-0.8em\TeX}}}

\setcopyright{acmcopyright}
\copyrightyear{2023}
\acmYear{2023}
\acmDOI{xx.xxxx/xxxxxxx.xxxxxxx}

\acmConference[WWW '23]{}
  {May 1--5, 2023}{Austin, Texas, USA}
\acmBooktitle{Proceedings of The Web Conference 2023 (WWW ’23), May 1–5, 2023, Austin, Texas, USA.}
\acmPrice{15.00}
\acmISBN{978-1-4503-XXXX-X/18/06}

\usepackage{enumitem}
\usepackage{multirow}
\newlist{questions}{enumerate}{2}
\setlist[questions,1]{label=\textbf{RQ\arabic*}.,ref=\textbf{RQ\arabic*}}
\setlist[questions,2]{label=\textbf{(\alph*)},ref=\textbf{\thequestionsi(\alph*)}}

\begin{document}

\title{How Many Tweets Do We Need?: Efficient Mining of Short-Term Polarized Topics on Twitter: A Case Study From Japan}

\author{Tomoki Fukuma}
\email{tomoki.fukuma@tdailab.com}
\orcid{1234-5678-9012}
\authornotemark[1]
\affiliation{%
  \institution{TDAI Lab Co.,Ltd.}
}

\author{Koki Noda}
\affiliation{%
  \institution{TDAI Lab Co.,Ltd.}
}
\email{koki.noda@tdailab.com}

\author{Hiroki Kumagai}
\affiliation{%
  \institution{TDAI Lab Co.,Ltd.}
}
\email{hiroki.kumagai@tdailab.com}

\author{Hiroki Yamamoto}
\affiliation{%
  \institution{TDAI Lab Co.,Ltd.}
}
\email{hiroki.yamamoto@tdailab.com}

\author{Yoshiharu Ichikawa}
\affiliation{%
  \institution{NHK (Japan Broadcasting Corporation)}
}
\email{ichikawa.y-gq@nhk.or.jp}

\author{Kyosuke Kambe}
\affiliation{%
  \institution{NHK (Japan Broadcasting Corporation)}
}
\email{kambe.k-je@nhk.or.jp}

\author{Yu Maubuchi}
\affiliation{%
  \institution{NHK (Japan Broadcasting Corporation)}
}
\email{masubuchi.y-lq@nhk.or.jp}

\author{Fujio Toriumi}
\affiliation{%
  \institution{The University of Tokyo}
}
\email{tori@sys.t.u-tokyo.ac.jp}


\begin{abstract}
In recent years, social media has been criticized for yielding polarization.
Identifying emerging disagreements and growing polarization is important for journalists to create alerts and provide more balanced coverage. 
While recent studies have shown the existence of polarization on social media, they primarily focused on limited topics such as politics with a large volume of data collected in the long term, especially over months or years. While these findings are helpful, they are too late to create an alert immediately. To address this gap, we develop a domain-agnostic mining method to identify polarized topics on Twitter in a short-term period, namely 12 hours. As a result, we find that daily Japanese news-related topics in early 2022 were polarized by 31.6\% within a 12-hour range. We also analyzed that they tend to construct information diffusion networks with a relatively high average degree, and half of the tweets are created by a relatively small number of people. However, it is very costly and impractical to collect a large volume of tweets daily on many topics and monitor the polarization due to the limitations of the Twitter API.
To make it more cost-efficient, we also develop a prediction method using machine learning techniques to estimate the polarization level using randomly collected tweets leveraging the network information. Extensive experiments show a significant saving in collection costs compared to baseline methods. In particular, our approach achieves F-score of 0.85, requiring 4,000 tweets, 4x savings than the baseline. To the best of our knowledge, our work is the first to predict the polarization level of the topics with low-resource tweets. Our findings have profound implications for the news media, allowing journalists to detect and disseminate polarizing information quickly and efficiently.

\end{abstract}





\maketitle


\section{Introduction}
The advent of social media had a profound impact on society, redefining how we obtain information. The traditional one-to-many information spreading has shifted to many-to-many communication, as anyone can easily speak up and share and choose whom to follow and with whom to interact. However, this decentralized nature of social media
encourages users to interact with information mostly aligned with their beliefs\cite{Himelboim2013BirdsOA} and has been criticized for fostering filter bubbles\cite{10.5555/2029079} and political polarization\cite{polarizationref1,doi:10.1177/2056305117729314}. A recent study\cite{polarfake} reported that misinformation might quickly propagate when the polarization level is high.

The rise of daily news is shaping distinct communication networks on social media.
Identifying the emerging disagreements and growing polarization on social media is crucial for journalists to create alerts and provide more balanced coverage. 
A recent research\cite{10.1145/3178876.3186130} suggests that by presenting social media users with a bird's eye view of an ideologically-fragmented social network and asking
them to identify their positions within it, can help cultivate intellectual humility and motivate more diverse content-sharing and information-seeking behaviors.

Though many studies have shown the existence of polarization on social media, prior studies are limited in two main ways. First, previous research mainly observed polarization on social networks collected for more than a month or years \cite{Garimella_Weber_2017, Conover_Ratkiewicz_Francisco_Goncalves_Menczer_Flammini_2021}. These findings are helpful but too late to create alerts immediately. To our knowledge, no studies have examined polarization in the short term, less than a day.
Second, prior research typically focused on a single or a minimal set of related topics, usually political issues. Various topics are debated daily, and polarization may occur on non-political subjects. Furthermore, most studies have been conducted in the United States context. The U.S. has a strong two-party system characterized by ideological and affective polarization along party lines. Such partisans are rarely seen in other countries.

Therefore, our study aims to better understand the nature of the polarized topic in the short term on Twitter on various topics. However, the expensive cost of data collection due to the limitations of the API hinders the analysis of a wide variety of topics every day. This limitation suggests the potential need for an efficient approach, which would require small subsets of tweets to predict the polarization level. Consequently, our work aims to answer three questions: (1) How often do polarized topics related to everyday news exist in a short-term period? To answer this question, we develop and employ a mining method to identify polarized topics on Twitter and conduct extensive research on originally collected datasets related to Japanese newsworthy events in early 2022. (2) What characteristics do polarized topics have compared to non-polarized topics? We compare the polarized and non-polarized topics from the perspective of the news genre, network size, average degree, URL ratio, Hashtag ratio, and vocal minority level, which how a small number of people created a large volume of content. (3) How many tweets do we need to collect to obtain reliable results on calculating polarization? Lastly, we assess whether predictions made from randomly chosen subsets of tweets align with the ground truth. Furthermore, we develop machine learning techniques incorporating network statistics, making better predictions with lower resources.
To the best of our knowledge, our work is the first to predict the polarization level of the topics with low-resource tweets.

Our contributions are as follows:
\begin{itemize}
    \item Our analysis shows that daily Japanese news-related topics in early 2022 were polarized by 31.6\% within a 12-hour range. 
    
    \item  We show that polarized topics tend to happen related to the news genre of $International$, $Business$, and $Social \cdot Culture$ and often create the large size of the information diffusion networks with a high average degree. Another interesting characteristic is that half of the tweets are created by a relatively small number of people.

    \item We implement a machine learning approach to estimate the polarization level using randomly collected tweets. Our analysis achieves F-score of 0.85, requiring 4,000 tweets, 4x savings than the baseline. 

\end{itemize}

\section{Related Work}\label{rw}

\subsection{Taxonomy of Polarization Detection Methods}
As news consumption shifted to social media, researchers began investigating how news content sharing habits on social media platforms are polarized. 
From a sociological point of view\cite{10.2307/2782461}, polarization is understood as the division of individuals into coherent and strongly opposed groups based on their opinions on one or more issues. Therefore previous studies can be classified according to how groups are defined and how dissimilarities of opinions among them are defined.

First, there are mainly two approaches for determining the groups: manual partitioning or network-based partitioning.
Manual partitioning means classifying users based on the analyst's intent, in most cases, political leaning. They are often used by predefined accounts \cite{Conover_Ratkiewicz_Francisco_Goncalves_Menczer_Flammini_2021,Garimella_Weber_2017}, use of predefined hashtags \cite{Rizoiu_Graham_Zhang_Zhang_Ackland_Xie_2018}, or news URLs\cite{baly-etal-2019-multi}. The other approach is network-based methods, utilizing graph partitioning methods\cite{10.1145/3140565} on retweet or mention networks to obtain groups. This type of method doesn't require prior knowledge and extracts groups rather than intentionally obtaining groups such as left or right.

Second, there are also mainly two approaches for determining the similarities of the opinions between the aforementioned groups: network-based and content based. Network-based approaches leverage information diffusion to inform ideological alignment. These models assume that users interact more with people of similar opinions. Interactions can be retweets \cite{7454756, Morales2015MeasuringPP}, followings \cite{Barbera}, mentions\cite{Conover_Ratkiewicz_Francisco_Goncalves_Menczer_Flammini_2021} or multi-relational network (combination of retweets, mentions, likes and follows)\cite{10.1145/3394486.3403275}. A previous study reported those network-based text clustering potentially yields better results rather than topic modeling\cite{uchida}.  
Content-based approaches leverage information related to users' tweets and other textual data\cite{He2021DetectingPT}. Prior research mainly leveraged for: hashtag\cite{Garimella_Weber_2017,6113114,10.1145/3140565}, lexicon dictionaries\cite{https://doi.org/10.48550/arxiv.1409.8152}, sentiment\cite{Mejova2014ControversyAS}, stance\cite{8181508}, word embedding\cite{preotiuc-pietro-etal-2017-beyond}.

The most well-known strategy for observing polarization is grouping users manually, namely left or right users, and seeing the disconnectivity between them in the social network. In our study, we group users based on network-based interactions since it is impossible to manually select opposing groups of users without prior domain-specific knowledge of every topic. Then we also use network-based disconnectivity to measure the dissimilarities of opinions. Previous content-based works are short-handed because URLs or hashtags are partially included. Also, using the stance detection method is difficult because we do not know the target people agree with beforehand. Our approach shares the same selection of methods and motivation in \cite{10.1145/3140565}.

\subsection{Data Collection on Polarization Detection}\label{sec2.2}
Recent studies mainly investigated polarization in the political domain, with case studies focusing on major, long-term events such as presidential elections. Therefore, in terms of data collection, choosing the data collection period has been event-driven and dependent on analysts' interests. Conover et al.\cite{Conover_Ratkiewicz_Francisco_Goncalves_Menczer_Flammini_2021} collected six-week tweets related to the 2010 U.S. congressional midterm elections. Mejova et al.\cite{https://doi.org/10.48550/arxiv.1409.8152} consider discussion of controversial and non-controversial news over seven months. This paper aims to identify and quantify short-term controversies on any topic discussed on social media rather than on a limited set of topics as in prior studies. Therefore, our analysis differs from most studies in limiting the collection period rather than the topic.

Those previous studies can be categorized into two according to their types of data collection. The first type collects tweets or user interaction information based on manually selected seed accounts, such as politicians, influencers, and media accounts\cite{Colleoni,barbera_2015,10.2307/45094371,HALBERSTAM201673}. Utilizing their account information, they collect past tweets or follower lists. While these methods can collect a wide range of relevant topics, there are difficulties in determining the accounts to be used for the survey. The second type collects tweets by using keywords or hashtag search. These studies manually select hashtags related to topics interested in \cite{Conover_Ratkiewicz_Francisco_Goncalves_Menczer_Flammini_2021,10.1145/3140565}, or use keywords in their tweets\cite{Himelboim2013BirdsOA}. While these methods make it easier to collect specific topics, they also tend to make it challenging to set keywords when analyzing broad topics such as political divisions. In our study, we employ the second type of method because our research targets polarization in a small discussion of specific news rather than a broader topic such as politics.

\begin{figure*}
    \centering
    \includegraphics[width=\linewidth]{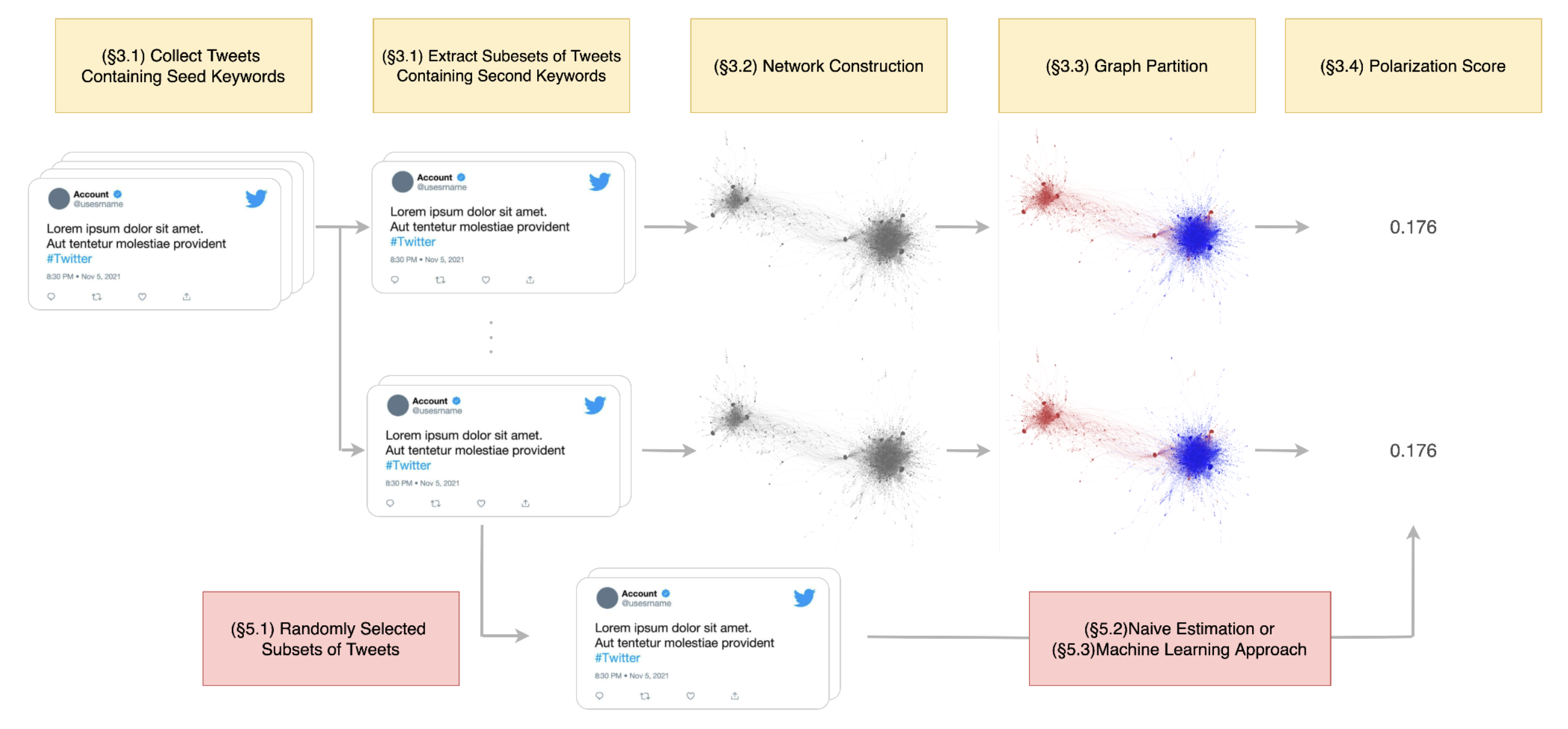}
    \caption{Overview of our polarized topic detection.}
    \label{fig:overview}
\end{figure*}

\section{Method for Identifying Polarized Topics}\label{methodology}
This section describes the method used to identify polarized topics from collected tweets. The overall framework in this paper is illustrated in Figure \ref{fig:overview}. 

\subsection{Data collection}\label{netconst}
To examine the spread of information on a single topic, we use tweets(posts on Twitter) to create a network. We collect tweets by using keywords as described in Section \ref{sec2.2} because we want to make an analysis on a narrow discussion related to specific news rather than a broader topic such as politics. However, discovering the search queries to find polarized topics by trial and error is difficult and time-consuming. 

To address this limitation, we employ a two-stage data collection method: First, we collect tweets with a single word ($seed$ keyword) that involves a variety of topics (e.g., "Ukraine"). Then, we extract subsets of tweets based on the smaller topics discussed. Topic models such as LDA\cite{10.5555/944919.944937} are one popular machine-learning approach to identifying topics. However, it requires the number of topics beforehand, and it is known that a too large or too small number of topics will affect the inference process and cause inaccuracies in grouping topics in the training model\cite{pmlr-v32-tang14}. Therefore, we improve the basic approach by considering the word's popularity. Specifically, we compute the frequency of all terms, and we retrieve the top-$k$ (we use $k=10$) $sub-$keywords from a given seed (e.g., "Kyiv" given "Ukraine" as seed). Then we extract subsets of tweets containing both seed and sub-keywords.  
In our experiments, we use MeCab\cite{kudo-etal-2004-applying} for word segmentation and POS tagging to select nouns for sub-keywords.

\subsection{Network Construction}\label{subnetwork}
For each dataset queried by seed and sub-keyword, we use retweet information to build the interaction network $G$. Specifically, we use retweets as a proxy for influence\cite{Morales2015MeasuringPP} and build directed network $G = (V, E) $. Whenever a user $u_i \in V$ retweets a message originally posted by user $u_j$, we assume that $u_i$ is influenced by $ u_j$'s ideas. Hence, a new directed link (($u_j \rightarrow u_i$) $\in E$) is created.

\subsection{Graph Partition}\label{graphpart}

Next, we split the network, aiming to obtain two groups of similar size with the fewest edges between them. We use the graph partitioning setting used in \cite{CHEN2021102348}: we use METIS\cite{doi:10.1137/S1064827595287997} algorithm, setting a maximum imbalance constraint of 3:7 on the partitioning algorithm since forcing the partitions to be perfectly balanced when real groups are not the same size will result in the partition to divide the larger group, which incorrectly inflates inter-group agreement relative to a within-group agreement. The METIS algorithm has a straightforward and intuitive interpretation and has been shown to work well with retweet networks\cite{10.1145/3140565}.

\subsection{Measuring Polarization}

Next, we measure how two groups obtained in Section \ref{graphpart} are weakly connected. A recent study \cite{10.1145/3512962} found well commonly-used structural polarization measures yield high polarization scores even for random networks with similar density and degree distributions networks, and the choice of measurement method is not as critical as normalization. Followed by the suggestions by \cite{10.1145/3512962}, we use Adaptive E-I Index\cite{CHEN2021102348} for measuring polarization which is easy to interpret and employ the normalization technique proposed by \cite{10.1145/3512962}.

\textbf{Adaptive E-I Index.} 
We calculate our polarization score $\Phi(G)$ as:

\begin{equation}
\Phi(G)=\frac{\sigma_{A A}+\sigma_{B B}-\left(\sigma_{A B}+\sigma_{B A}\right)}{\sigma_{A A}+\sigma_{B B}+\left(\sigma_{A B}+\sigma_{B A}\right)}
\end{equation}
where $\sigma_{A A}$ is the ratio of links within the community $A$ (similarly for $\sigma_{B B}$) and  $\sigma_{A B}$ is the observed number of links between the communities $A$ and $B$ divided by the number of all potential links. This measure is an extension of the E-I Index\cite{10.2307/2786835}, as it accounts for different community sizes by using the density of links within each community. When the two groups are the same size, $\Phi(G)$ reduces to the E-I Index\cite{ctx31826023050006531}.

\textbf{Normalization.} 
We calculate our normalized polarization score $\hat{\Phi}(G)$ as:
\begin{equation}
\hat{\Phi}(G)=\Phi(G)-\langle\Phi(G_{C M})\rangle
\end{equation}
where $\Phi(G)$ is the polarization score of the observed network and $\left\langle\Phi\left(G_{C M}\right)\right\rangle$ is the expected polarization score of graphs calculated for multiple instances of the randomized network fixing the number of nodes and degree-degree sequences\cite{10.1145/1159913.1159930}. This score corrects for the expected effect of the size and degree distribution of the network. Each randomized network is repartitioned before computing its polarization score by applying the same graph partitioning algorithm for the original network. We sampled the randomized networks 50 times and calculated the score.

\section{Polarization on Everyday, Short-term Topics}
In this section, we conduct experiments with the aim of answering the following research question using the methodology in Section \ref{methodology}: \textbf{RQ1. How often do polarized topics exist in everyday news topics?}

\begin{figure}
    \centering
    \includegraphics[width=\linewidth]{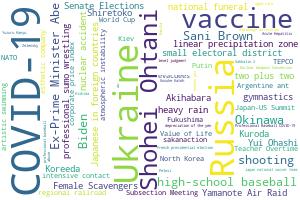}
    \caption{Word cloud visualization of seed keywords translated from Japanese used in our experiments. We select 100 seed keywords using KMeans clustering on 17,620 news articles collected from February 1 to August 4. The size of the words indicates the corresponding cluster size.}
    \label{fig:tsne}
\end{figure}

\subsection{Dataset Collection}\label{datacollection}
Since this work aims to create polarization alerts in a short-term period for journalists, we tried to collect a variety of topics related to daily news topics. Specifically, we collect tweets and create networks in the following steps.

First, in order to choose the seed keywords described in Section \ref{subnetwork}, we collect all the news from NHK news web, Japanese public broadcast media, from February 1 to August 4, 2022, to know the newsworthy events that happened in Japan. The overall dataset contains 17,620 news. These news articles have distinct categories, namely $International $, $Politics$, $Business$, $Social$, $Sports$, $Science \cdot Culture$, $Life$, $Weather \cdot Disaster$ and about 96 news items are published per day.

Second, to reduce the volume of collecting tweets, we extract 100 news articles and their seed keywords, keeping the diversity of the topics. In particular, we convert each news title into 43,590-dimensional Bag-of-Words vectors. Then we conduct KMeans clustering and select the 100 articles closest to their centers. After obtaining articles, we manually select seed keywords which we may include a relatively broad range of topics related to that news. Figure \ref{fig:tsne} illustrates the word cloud visualization of seed keywords translated from Japanese, and the size indicates the corresponding cluster size, where the big size of the words suggests that there are large amount of news related to that seed keywords.

\begin{figure}
    \centering
    \includegraphics[width=\linewidth]{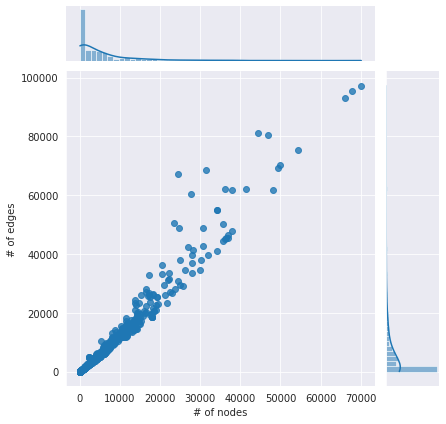}
    \caption{Scatter plot showing number of nodes and edges of the 1,000 retweet networks obtained in Section \ref{datacollection}}
    \label{fig:nodeedgescatter}
\end{figure}

Finally, Twitter API v2 for Academic Research was used to collect tweets containing the seed keywords. We collect tweets in a period of 12 hours after the corresponding news was published. As mentioned in Section \ref{subnetwork}, after collecting tweets, we extract top-$10$ sub-keywords for each dataset based on frequency, which results in 1,000 sub-topics. However, excluding sub-topics that were too small to create a network, a total of 842 networks were created. Network statistics are presented in Figure \ref{fig:nodeedgescatter}.

\subsection{Distributions of Polarization}

\begin{figure}
    \centering
    \includegraphics[width=\linewidth]{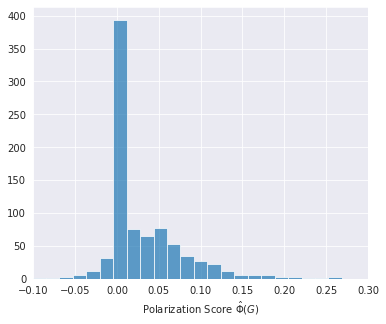}
    \caption{Histogram of polarization scores $\hat{\Phi}(G)$ calculated from 842 networks related to Japanse news from February to July 2022}
    \label{fig:pairplot}
\end{figure}

\begin{figure*}
    \centering
    \includegraphics[width=.8\linewidth]{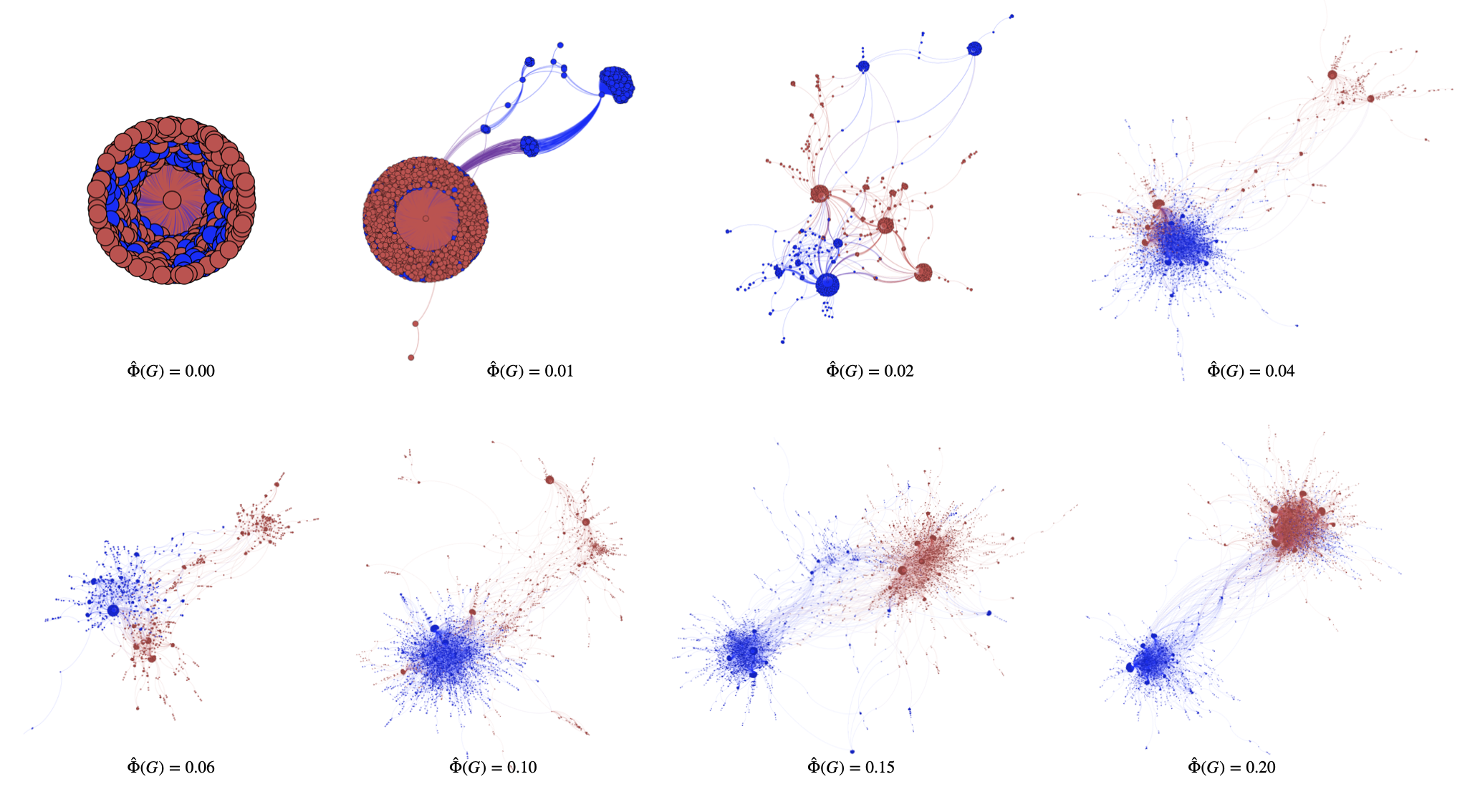}
    \caption{Sample retweet network visualized using the ForceAtlas2 algorithm with calculated polarization scores $\hat{\Phi}(G)$.}
    \label{fig:sample}
\end{figure*}

Here we investigate to what extent the polarization happens related to the Japanese news event in 12 hours. For this analysis, we compute polarization scores for each network. Figure \ref{fig:pairplot} shows the histogram of the polarization scores $\hat{\Phi}(G)$ over the 842 networks we collected. Inspecting this plot, we find that 39.9\% of the networks have a polarization score of 0 or less.
We show sample networks according to polarization score in Figure \ref{fig:sample}. The partitions are drawn in blue or red. The graph layouts are generated by ForceAtlas2 algorithm\cite{fa2} and are based solely on the structure of the graph, not on the partitioning computed by METIS. 
As shown in Figure \ref{fig:sample}, topics in which a small number of tweets are largely retweeted show low polarization scores because the structure changes little after shuffling.
By manual inspection, we use a threshold of $0.04$ to binarize the networks into polarized or not, and we found that 31.6\% of all the networks are polarized.



\subsection{Characterizing Polarization Topic}\label{sec:chara}
This subsection aims to answer the following research question: \textbf{RQ2. What are the characteristics of polarized topics compared to non-polarized ones?}

To get a more detailed view of the characteristics that are related to polarization, we compare how the networks of polarized and non-polarized topics differ. To characterize the difference, we use the following viewpoints: news genre, network size, the ratio of the use of hashtags and URLs, and vocal minority.
As in the previous section, we use 0.04 as the threshold to determine polarization. 

\begin{figure*}
    \centering
    \includegraphics[width=\linewidth]{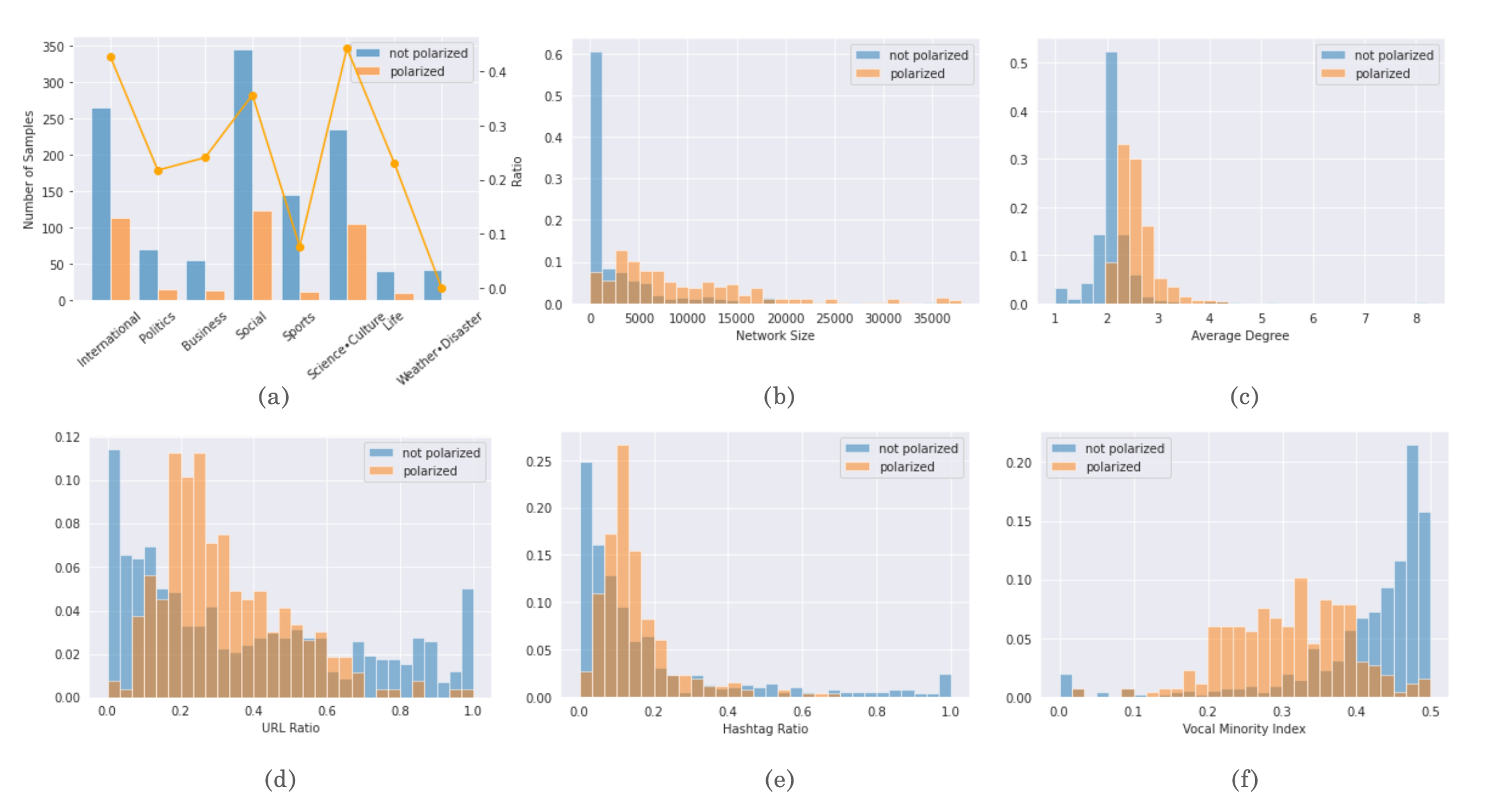}
    \caption{Comparison of the characteristics of the network between polarized and non-polarized topics:(a) news genre, (b) network size, (c) average degree, (d) (e) ratio of the use of Hashtags and URLs, and (f) vocal minority.}
    \label{fig:chara}
\end{figure*}

\textbf{Genre: }
First, we connect polarization with the news genre from the original news data. Figure \ref{fig:chara} (a) displays a bar plot of the network size and a line plot showing the ratio of polarization across different genres. Our results indicate that in comparison with the rest of the topics, $International$, $Business$, and $Social$ are among the more polarized. We found that polarization does not occur at such a high rate on political topics, which have been extensively analyzed in previous studies. Our results suggest that more analysis in broader genres is needed to create polarization alerts.

\textbf{Network size: }
We also analyze the relationship between the network size and polarization. This comparison is shown in Figure \ref{fig:chara} (b). The pattern is very apparent: as network size increases, the polarization ratio goes up. This is because the smaller size of the network results in a smaller difference from the randomly shuffled network in the calculation of polarization.

\textbf{Average Degree:}
The average degree is simply the average number of edges per node in the graph. The higher the value indicates, the more dense the network. This comparison is shown in Figure \ref{fig:chara} (c). We can see the dense information diffusion network are more easily polarized. This shows that networks that have many mutual retweet relationships are more likely to be polarized than in a situation where a few tweets are spread out in large numbers.

\textbf{URLs \& Hashtags: }
Next, we comment on the use of URLs and hashtags contained in tweets. This comparison is shown in Figure \ref{fig:chara} (d) (e). These results show that polarization rarely occurs in topics with URL ratio as low as around 0.1 or as high as 0.8 or higher, and polarization tends to happen in the rest of the in-between areas. In terms of Hashtag ratio, we can see that the polarized topics relatively include more use of Hashtag. 

\textbf{Vocal Minority:}
Another interesting metric for analysis is the vocal minority, the small group of individuals that frequently and strongly voice their opinions. Previous study\cite{10.1145/3209581} explored the Facebook dataset from 2009 with almost 40,000 active users and found 7\% of them produced 50\% of the posts.
Inspired by this analysis, we introduce the vocal minority index, what percentage of the active user created the 50\% of the tweets. We show these statistics in Figure \ref{fig:chara} (f). Our results is very apparent: as vocal minority level increases, ratio of polarization goes up.


\section{Efficient Estimation of Polarization}
As seen in the previous section, we confirmed that polarization happens daily. However, the expensive cost of data collection due to the limit of the API hinders the analysis of polarization on wide variety of topics every day. This limitation suggests the potential need for an efficient approach, which would require small
subsets of tweets to predict the polarization level. To the best of our knowledge, our work is the first to predict the polarization level of the topics with low-resource tweets. Therefore this section aims to answer the following question: \textbf{RQ3. How many tweets do we need to collect to obtain reliable results on the calculation of polarization?}

    
\begin{figure}
    \centering
    \includegraphics[width=\linewidth]{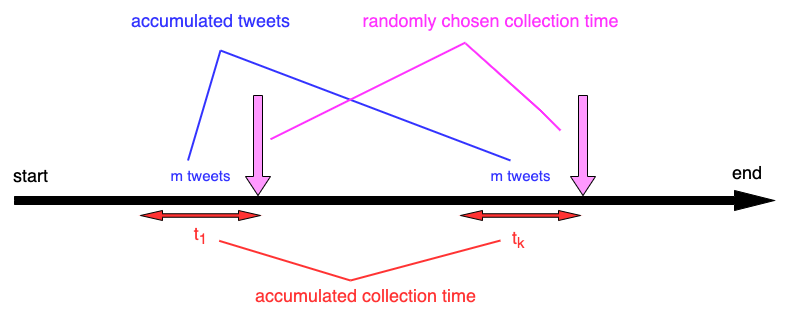}
    \caption{Data sampling procedure. We randomly select $k$ points from the considering period and pick a subset of $m$ tweets.}
    \label{fig:mech}
\end{figure}

\subsection{Experimental Setting}
We now assess whether predictions made from randomly chosen subsets of tweets align with the ground truth. Figure \ref{fig:mech} shows the data sampling procedure. First, we randomly select $k$ points as a start of tweet collection from the considering period. Then we collect $m$ tweets back from each point. In this work, we use $m=100$, the maximum capacity for collecting tweets with one request. Note that reducing the $m$ leads to an increase in API executions. 

We implement two variations of estimation methods described in Section \ref{naive} and Section \ref{mlapp} using the randomly chosen subsets of tweets. Then we evaluate the estimation results with the scores obtained from the whole dataset, which constitutes the ground truth. We evaluate performance in terms of $R^2$ score, precision, and recall. We use k=10, 20, 40, 80, 160, and 320 and report the average scores of 10 trials. The overall framework is also illustrated in Figure \ref{fig:overview}.

\subsection{Results of Naive Estimation}\label{naive}

\begin{figure*}
    \centering
    \includegraphics[width=\linewidth]{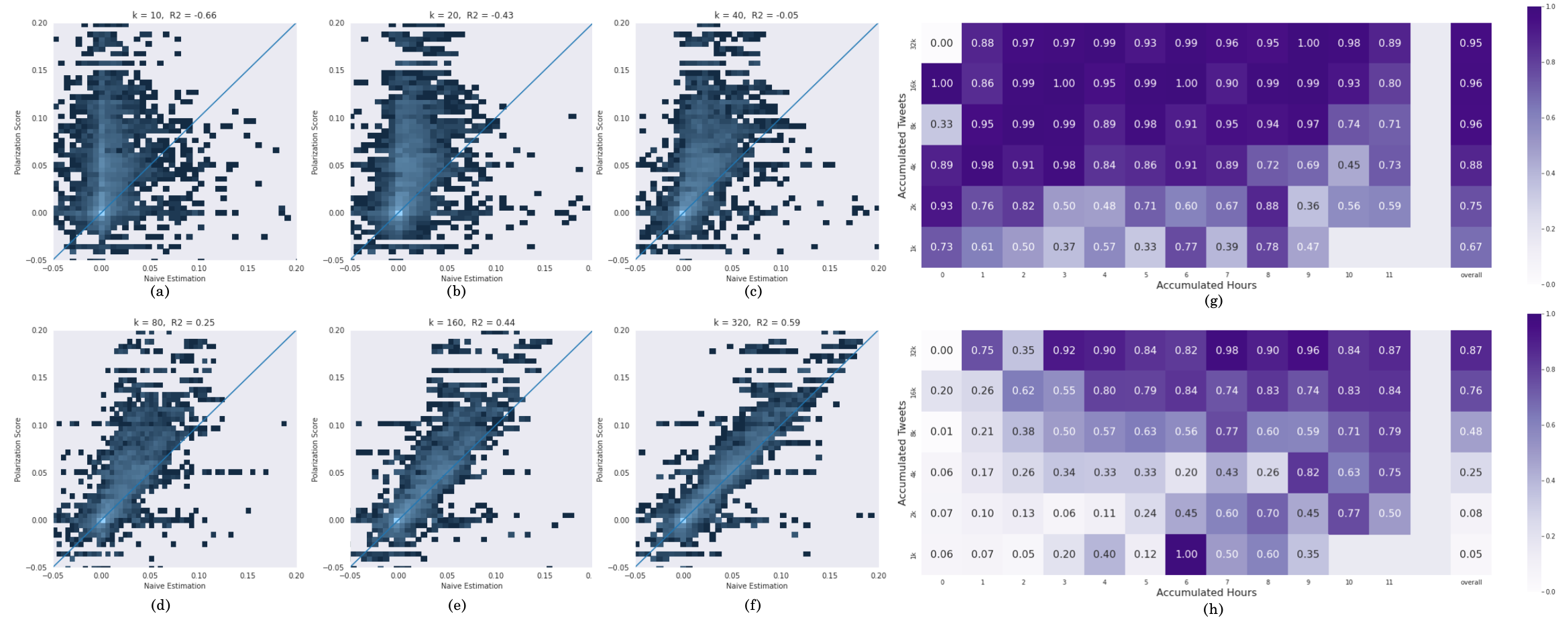}
    \caption{ Results of the naive estimation: (a) $\sim$ (f) shows the scatter density plot of the estimation and the ground truth, and (g) (h) shows precision and recall according to the collected amount of tweets and accumulated hour-levels of the tweets.}
    \label{fig:res1}
\end{figure*}

Naive estimation calculates the polarization score from the network constructed from obtained subsets of tweets using the method described in Section \ref{methodology}. 
Results obtained for each dataset are shown in Figure \ref{fig:res1} (a) $\sim$ (f) showing the scatter density plot of the prediction(x-axis) and the ground truth(y-axis). Figure \ref{fig:res1} (g) (h) shows the precision and recall with the x-axis showing accumulated hour-levels of the collection and the y-axis showing the depending on the number of collected points. The last column in Figure \ref{fig:res1} (g) (h) shows the overall results of precision and recall.
This evaluation aims to understand the performance according to how many tweets are collected and how many hours they occupy, which can estimate the original size of the networks. A single pixel corresponds to multiple networks, showing the average performance. Thus the number of networks in pixels horizontally from 0 hour-level to 11 hour-level is summed up to 842 for all rows. Blanks are provided where no network exists. 

As shown in Figure \ref{fig:res1} (a) $\sim$ (f), we can see a clear trend that the smaller the number of tweets collected, the lower the predicted polarization score. This insight can help understand the results in Figure \ref{fig:res1} (g) (h), showing that precision is relatively high while recall tends to be low. Another interesting finding is that Figure \ref{fig:res1} (g) produces low performance in the right-bottom corner, which suggests that the small polarized networks are hard to predict even if they are collected most of the time. At the same time, it produces 0.73 of precision to the networks by collecting just an hour of tweets.

\begin{figure*}
    \centering
    \includegraphics[width=\linewidth]{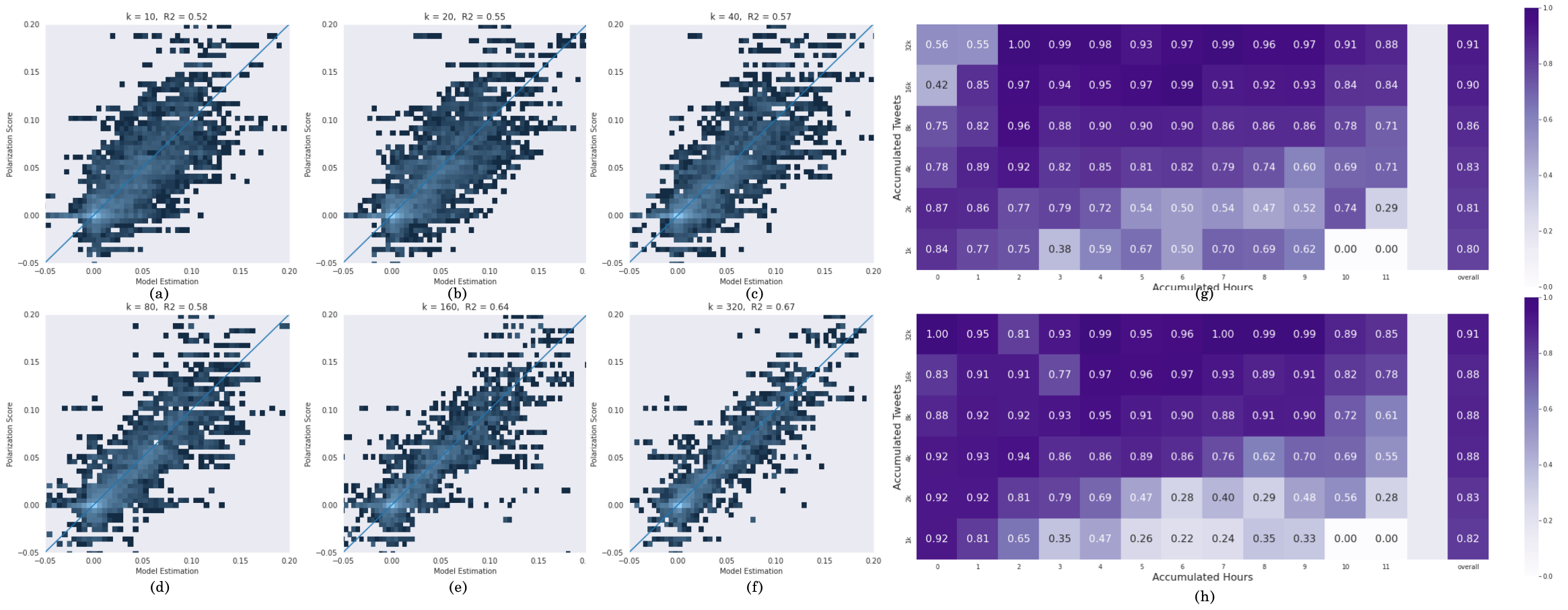}
    \caption{Results of the machine learning approach same as in Figure \ref{fig:res1}: (a) $\sim$ (f) shows the scatter density plot of the estimation and the ground truth, and (g) (h) shows precision and recall according to the collected amount of tweets and accumulated hour-levels of the tweets. }
    \label{fig:stra2}
\end{figure*}

\subsection{Results of Machine Learning Approach}\label{mlapp}
Up to this point, we are very interested in whether a prediction can be better by leveraging network or user information seen in Section \ref{sec:chara}.
In this subsection, we will evaluate the effectiveness of the combination of the features for predicting polarization scores. We use Random Forest\cite{rf} to predict the polarization score $\hat{\Phi}(G)$ using features, news genre, network size, URL ratio, Hashtag Ratio, and Vocal Minority Index. The training uses a 10-fold cross-validation procedure. The training dataset is split into ten smaller sets, and the model is trained using nine folds and validated on the remaining part of the data. To evaluate the performance of the trained model, we report the same metric in Section \ref{naive}. Note that we binarize the network into polarized or not polarized after making predictions using the same threshold in Section \ref{sec:chara} to calculate precision and recall.

Figure \ref{fig:stra2} shows the results of the machine learning approach. Comparing the results in Figure \ref{fig:stra2} (a) $\sim$ (f) and Figure \ref{fig:res1} (a) $\sim$ (f), our approach achieves F-score of 0.85(precision of  0.83 and recall of 0.88), requiring 4,000 tweets, 4x savings than the baseline(precision of  0.96 and recall of 0.76). We can also see significant improvement in recall where collecting just 1,000 tweets can produce rivaling results with the naive estimation results with 16,000 $\sim$ 32,000 tweets. In Figure \ref{fig:stra2} (g) (h) and  Figure \ref{fig:res1} (g) (h), we can see large improvement of recall especially in left part of the matrix (large networks) from naive estimation.  Our model consistently shows a similar trend: it produces low precision and recall in the lower right, which suggests that small polarized networks are hard to detect with small data. Our results conclude that it is more beneficial to leverage knowledge from another relevant domain when there is less data. 

\section{Conclusion}

Identifying emerging disagreements and growing polarization is crucial for journalists to create alerts and provide more balanced coverage. In this work, we studied the efficient detection of polarized topics related to newsworthy Japanese events. Our study revealed that polarization occurs in everyday topics, regardless of whether they are political or not. Moreover, polarization is often accompanied by voice minority and contains a relatively high ratio of bot activities. Last but not least, incorporating those features can predict polarization even with a randomly selected small amount of tweets. We have shown that the machine learning approach achieves F-score of 0.85, requiring 4,000 tweets, 4x savings than the baseline. 

One possible thing to do in the future is to optimize the strategy of collecting tweets. Although we attempted to extract tweets randomly this time, better predictions may be possible by controlling the next selection of collection time depending on how many minutes were collected. This notion is based on the idea that there may be a segment of time that is important for collecting tweets since the burst in tweets is not constant and may be concentrated at a certain time of the day. Another possible future approach is investigating more features to predict prediction and consideration of different machine learning methods. In this study, we focused on the basic feature of the network and the frequency of statements. However, there may be unexplored features such as followers' information and the semantic alignment of the tweets.

Our work is the first to show that short-term polarization occurs in a short period on any topic and that they are predictable with a small number of tweets. These results have profound implications for the use of news media. This allows journalists to detect and disseminate polarizing information quickly. Prior studies have proposed many techniques to eliminate polarization, suggesting new recommendation algorithms on social media. However, changing the recommendation system involves great cost and risk. Our study is a step toward eliminating polarization for everyone in that we are doing what we can as a media outlet. While some research has shown the backfire effects\cite{Nyhan2010}, strengthening of polarization through the presentation of diverse opinions. However, not enough is known about what happens when these effects occur on the other topics, similar to the purpose of this study. In the future, verifying the effectiveness of presenting the various opinions obtained from these studies to users will be necessary.

\bibliographystyle{ACM-Reference-Format}
\bibliography{sample-base}

@article{10.1145/3512962,
author = {Salloum, Ali and Chen, Ted Hsuan Yun and Kivel\"{a}, Mikko},
title = {Separating Polarization from Noise: Comparison and Normalization of Structural Polarization Measures},
year = {2022},
issue_date = {April 2022},
publisher = {Association for Computing Machinery},
address = {New York, NY, USA},
volume = {6},
number = {CSCW1},
url = {https://doi.org/10.1145/3512962},
doi = {10.1145/3512962},
journal = {Proc. ACM Hum.-Comput. Interact.},
month = {apr},
articleno = {115},
numpages = {33},
keywords = {polarization, clustering, sociology, network science, normalization, computational social science, statistical significance, networks, twitter, political polarization, community detection}
}

@inproceedings{10.1145/1134271.1134277,
author = {Adamic, Lada A. and Glance, Natalie},
title = {The Political Blogosphere and the 2004 U.S. Election: Divided They Blog},
year = {2005},
isbn = {1595932151},
publisher = {Association for Computing Machinery},
address = {New York, NY, USA},
url = {https://doi.org/10.1145/1134271.1134277},
doi = {10.1145/1134271.1134277},
booktitle = {Proceedings of the 3rd International Workshop on Link Discovery},
pages = {36–43},
numpages = {8},
keywords = {social networks, political blogs, link analysis},
location = {Chicago, Illinois},
series = {LinkKDD '05}
}

@article{Conover_Ratkiewicz_Francisco_Goncalves_Menczer_Flammini_2021, title={Political Polarization on Twitter}, volume={5}, url={https://ojs.aaai.org/index.php/ICWSM/article/view/14126}, abstractNote={ &lt;p&gt; In this study we investigate how social media shape the networked public sphere and facilitate communication between communities with different political orientations. We examine two networks of political communication on Twitter, comprised of more than 250,000 tweets from the six weeks leading up to the 2010 U.S. congressional midterm elections. Using a combination of network clustering algorithms and manually-annotated data we demonstrate that the network of political retweets exhibits a highly segregated partisan structure, with extremely limited connectivity between left- and right-leaning users. Surprisingly this is not the case for the user-to-user mention network, which is dominated by a single politically heterogeneous cluster of users in which ideologically-opposed individuals interact at a much higher rate compared to the network of retweets. To explain the distinct topologies of the retweet and mention networks we conjecture that politically motivated individuals provoke interaction by injecting partisan content into information streams whose primary audience consists of ideologically-opposed users. We conclude with statistical evidence in support of this hypothesis. &lt;/p&gt; }, number={1}, journal={Proceedings of the International AAAI Conference on Web and Social Media}, author={Conover, Michael and Ratkiewicz, Jacob and Francisco, Matthew and Goncalves, Bruno and Menczer, Filippo and Flammini, Alessandro}, year={2021}, month={Aug.}, pages={89-96} }

@article{10.1145/3140565,
author = {Garimella, Kiran and Morales, Gianmarco De Francisci and Gionis, Aristides and Mathioudakis, Michael},
title = {Quantifying Controversy on Social Media},
year = {2018},
issue_date = {March 2018},
publisher = {Association for Computing Machinery},
address = {New York, NY, USA},
volume = {1},
number = {1},
issn = {2469-7818},
url = {https://doi.org/10.1145/3140565},
doi = {10.1145/3140565},
journal = {Trans. Soc. Comput.},
month = {jan},
articleno = {3},
numpages = {27},
keywords = {filter bubble, polarization, Controversy, echo chambers, twitter}
}

@article{Mejova2014ControversyAS,
  title={Controversy and Sentiment in Online News},
  author={Yelena Mejova and Amy X. Zhang and Nicholas A. Diakopoulos and Carlos Castillo},
  journal={ArXiv},
  year={2014},
  volume={abs/1409.8152}
}

@inproceedings{he2021detecting,
  title={Detecting Polarized Topics Using Partisanship-aware Contextualized Topic Embeddings},
  author={He, Zihao and Mokhberian, Negar and C{\^a}mara, Ant{\'o}nio and Abeliuk, Andres and Lerman, Kristina},
  booktitle={Findings of the Association for Computational Linguistics: EMNLP 2021},
  pages={2102--2118},
  year={2021}
}

@inproceedings{HighCorrelationbetweenIncoming,
author = {Saez-Trumper, Diego and Nettleton, David and Baeza-Yates, Ricardo},
year = {2011},
month = {01},
pages = {},
title = {High Correlation between Incoming and Outgoing Activity: A Distinctive Property of Online Social Networks?}
}

@inproceedings{10.1145/2660460.2660477,
author = {Saez-Trumper, Diego and Liu, Yabing and Baeza-Yates, Ricardo and Krishnamurthy, Balachander and Mislove, Alan},
title = {Beyond CPM and CPC: Determining the Value of Users on OSNs},
year = {2014},
isbn = {9781450331982},
publisher = {Association for Computing Machinery},
address = {New York, NY, USA},
url = {https://doi.org/10.1145/2660460.2660477},
doi = {10.1145/2660460.2660477},
booktitle = {Proceedings of the Second ACM Conference on Online Social Networks},
pages = {161–168},
numpages = {8},
keywords = {online advertising, online social networks, user value},
location = {Dublin, Ireland},
series = {COSN '14}
}

@inproceedings{10.1145/1772690.1772751,
author = {Kwak, Haewoon and Lee, Changhyun and Park, Hosung and Moon, Sue},
title = {What is Twitter, a Social Network or a News Media?},
year = {2010},
isbn = {9781605587998},
publisher = {Association for Computing Machinery},
address = {New York, NY, USA},
url = {https://doi.org/10.1145/1772690.1772751},
doi = {10.1145/1772690.1772751},
booktitle = {Proceedings of the 19th International Conference on World Wide Web},
pages = {591–600},
numpages = {10},
keywords = {homophily, pagerank, Twitter, influential, reciprocity, degree of separation, retweet, information diffusion, online social network},
location = {Raleigh, North Carolina, USA},
series = {WWW '10}
}

@inproceedings{10.1145/1298306.1298311,
author = {Mislove, Alan and Marcon, Massimiliano and Gummadi, Krishna P. and Druschel, Peter and Bhattacharjee, Bobby},
title = {Measurement and Analysis of Online Social Networks},
year = {2007},
isbn = {9781595939081},
publisher = {Association for Computing Machinery},
address = {New York, NY, USA},
url = {https://doi.org/10.1145/1298306.1298311},
doi = {10.1145/1298306.1298311},
booktitle = {Proceedings of the 7th ACM SIGCOMM Conference on Internet Measurement},
pages = {29–42},
numpages = {14},
keywords = {social networks, measurement, analysis},
location = {San Diego, California, USA},
series = {IMC '07}
}

@article{InfluenceandPassivityinSocialMedia,
author = {Romero, Daniel and Galuba, Wojciech and Asur, Sitaram and Huberman, Bernardo},
year = {2011},
month = {01},
pages = {1-9},
title = {Influence and Passivity in Social Media},
journal = {Inf. Syst. J.},
doi = {10.1145/1963192.1963250}
}

@inproceedings{10.1145/1963405.1963504,
author = {Wu, Shaomei and Hofman, Jake M. and Mason, Winter A. and Watts, Duncan J.},
title = {Who Says What to Whom on Twitter},
year = {2011},
isbn = {9781450306324},
publisher = {Association for Computing Machinery},
address = {New York, NY, USA},
url = {https://doi.org/10.1145/1963405.1963504},
doi = {10.1145/1963405.1963504},

booktitle = {Proceedings of the 20th International Conference on World Wide Web},
pages = {705–714},
numpages = {10},
keywords = {communications, information flow, classification, twitter, communication networks, two-step flow},
location = {Hyderabad, India},
series = {WWW '11}
}

@article{Garimella_Weber_2017, title={A Long-Term Analysis of Polarization on Twitter}, volume={11}, url={https://ojs.aaai.org/index.php/ICWSM/article/view/14918}, number={1}, journal={Proceedings of the International AAAI Conference on Web and Social Media}, author={Garimella, Venkata Rama Kiran and Weber, Ingmar}, year={2017}, month={May}, pages={528-531} }

@article{polarfake,
author = {Del Vicario, Michela and Quattrociocchi, Walter and Scala, Antonio and Zollo, Fabiana},
year = {2019},
month = {03},
pages = {},
title = {Polarization and Fake News: Early Warning of Potential Misinformation Targets},
volume = {13},
journal = {ACM Transactions on the Web},
doi = {10.1145/3316809}
}

@article{10.2307/2786835,
 ISSN = {01902725},
 URL = {http://www.jstor.org/stable/2786835},
 abstract = {This paper argues that organizations with a particular social network structure are more effective than most organizations in responding to crises. Further, it is argued that the effective structure does not occur naturally, but must be designed consciously and carefully. A theory is developed based on well-founded principles of social science, most notably work on formal structure, conflict, friendships, and organizational crises. The paper concludes with an experimental test of one of the four propositions deduced from the theory. Six trials of the experiment found significant support for this propositon.},
 author = {David Krackhardt and Robert N. Stern},
 journal = {Social Psychology Quarterly},
 number = {2},
 pages = {123--140},
 publisher = {[Sage Publications, Inc., American Sociological Association]},
 title = {Informal Networks and Organizational Crises: An Experimental Simulation},
 urldate = {2022-08-14},
 volume = {51},
 year = {1988}
}

@article{Morales2015MeasuringPP,
  title={Measuring Political Polarization: Twitter shows the two sides of Venezuela},
  author={Alfredo J. Morales and Javier Borondo and Juan Carlos Losada and Rosa M. Benito},
  journal={Chaos},
  year={2015},
  volume={25 3},
  pages={
          033114
        }
}







@article{doi:10.1137/S1064827595287997,
author = {Karypis, George and Kumar, Vipin},
title = {A Fast and High Quality Multilevel Scheme for Partitioning Irregular Graphs},
journal = {SIAM Journal on Scientific Computing},
volume = {20},
number = {1},
pages = {359-392},
year = {1998},
doi = {10.1137/S1064827595287997},

URL = { 
        https://doi.org/10.1137/S1064827595287997
    
},
eprint = { 
        https://doi.org/10.1137/S1064827595287997
    
}

}











@article{
doi:10.1126/science.aal4230,
author = {Aylin Caliskan  and Joanna J. Bryson  and Arvind Narayanan },
title = {Semantics derived automatically from language corpora contain human-like biases},
journal = {Science},
volume = {356},
number = {6334},
pages = {183-186},
year = {2017},
doi = {10.1126/science.aal4230},
URL = {https://www.science.org/doi/abs/10.1126/science.aal4230},
eprint = {https://www.science.org/doi/pdf/10.1126/science.aal4230},
abstract = {AlphaGo has demonstrated that a machine can learn how to do things that people spend many years of concentrated study learning, and it can rapidly learn how to do them better than any human can. Caliskan et al. now show that machines can learn word associations from written texts and that these associations mirror those learned by humans, as measured by the Implicit Association Test (IAT) (see the Perspective by Greenwald). Why does this matter? Because the IAT has predictive value in uncovering the association between concepts, such as pleasantness and flowers or unpleasantness and insects. It can also tease out attitudes and beliefs—for example, associations between female names and family or male names and career. Such biases may not be expressed explicitly, yet they can prove influential in behavior. Science, this issue p. 183; see also p. 133 Computers can learn which words go together more or less often and can thus mimic human performance on a test of implicit bias. Machine learning is a means to derive artificial intelligence by discovering patterns in existing data. Here, we show that applying machine learning to ordinary human language results in human-like semantic biases. We replicated a spectrum of known biases, as measured by the Implicit Association Test, using a widely used, purely statistical machine-learning model trained on a standard corpus of text from the World Wide Web. Our results indicate that text corpora contain recoverable and accurate imprints of our historic biases, whether morally neutral as toward insects or flowers, problematic as toward race or gender, or even simply veridical, reflecting the status quo distribution of gender with respect to careers or first names. Our methods hold promise for identifying and addressing sources of bias in culture, including technology.}}

@article{polarizationref1,
author = {Rogowski, Jon and Sutherland, Joseph L.},
year = {2016},
month = {06},
pages = {},
title = {How Ideology Fuels Affective Polarization},
volume = {38},
journal = {Political Behavior},
doi = {10.1007/s11109-015-9323-7}
}




@article{doi:10.1177/2056305117729314,
author = {Jacob L. Nelson and James G. Webster},
title ={The Myth of Partisan Selective Exposure: A Portrait of the Online Political News Audience},
journal = {Social Media + Society},
volume = {3},
number = {3},
pages = {2056305117729314},
year = {2017},
doi = {10.1177/2056305117729314},

URL = { 
        https://doi.org/10.1177/2056305117729314
    
},
eprint = { 
        https://doi.org/10.1177/2056305117729314
    
}
,
    abstract = { Many assume that in a digital environment with a wide range of ideologically tinged news outlets, partisan selective exposure to like-minded speech is pervasive and a primary cause of political polarization. Yet, partisan selective exposure research tends to stem from experimental or self-reported data, which limits the applicability of their findings in a high-choice media environment. We explore observed online audience behavior data to present a portrait of the actual online political news audience. We find that this audience frequently navigates to news sites from Facebook, and that it congregates among a few popular, well-known political news sites. We also find that political news sites comprise ideologically diverse audiences, and that they share audiences with nearly all smaller, more ideologically extreme outlets. Our results call into question the strength of the so-called red/blue divide in actual web use. }
}



@book{10.5555/2029079,
author = {Pariser, Eli},
title = {The Filter Bubble: What the Internet Is Hiding from You},
year = {2011},
isbn = {1594203008},
publisher = {Penguin Group , The},
}

@Article{info:doi/10.2196/29570,
author="Jiang, Julie
and Ren, Xiang
and Ferrara, Emilio",
title="Social Media Polarization and Echo Chambers in the Context of COVID-19: Case Study",
journal="JMIRx Med",
year="2021",
month="Aug",
day="5",
volume="2",
number="3",
pages="e29570",
keywords="social media; opinion; infodemiology; infoveillance; COVID-19; case study; polarization; communication; Twitter; echo chamber",
issn="2563-6316",
doi="10.2196/29570",
url="https://med.jmirx.org/2021/3/e29570",
url="https://doi.org/10.2196/29570",
url="http://www.ncbi.nlm.nih.gov/pubmed/34459833"
}

@inproceedings{He2021DetectingPT,
  title={Detecting Polarized Topics Using Partisanship-aware Contextualized Topic Embeddings},
  author={Zihao He and Negar Mokhberian and Antonio Camara and Andr{\'e}s Abeliuk and Kristina Lerman},
  booktitle={EMNLP},
  year={2021}
}

@ARTICLE{7454756,  author={Wong, Felix Ming Fai and Tan, Chee Wei and Sen, Soumya and Chiang, Mung},  journal={IEEE Transactions on Knowledge and Data Engineering},   title={Quantifying Political Leaning from Tweets, Retweets, and Retweeters},   year={2016},  volume={28},  number={8},  pages={2158-2172},  doi={10.1109/TKDE.2016.2553667}}

@article{Barbera,
author = {Barberá, Pablo and Jost, John and Nagler, Jonathan and Tucker, Joshua and Bonneau, Richard},
year = {2015},
month = {08},
pages = {},
title = {Tweeting From Left to Right: Is Online Political Communication More Than an Echo Chamber?},
volume = {26},
journal = {Psychological science},
doi = {10.1177/0956797615594620}
}

@INPROCEEDINGS{6113114,  author={Conover, Michael D. and Goncalves, Bruno and Ratkiewicz, Jacob and Flammini, Alessandro and Menczer, Filippo},  booktitle={2011 IEEE Third International Conference on Privacy, Security, Risk and Trust and 2011 IEEE Third International Conference on Social Computing},   title={Predicting the Political Alignment of Twitter Users},   year={2011},  volume={},  number={},  pages={192-199},  doi={10.1109/PASSAT/SocialCom.2011.34}}

@inproceedings{Preotiuc,
author = {Preotiuc-Pietro, Daniel and Liu, Ye and Hopkins, Daniel and Ungar, Lyle},
year = {2017},
month = {01},
pages = {729-740},
title = {Beyond Binary Labels: Political Ideology Prediction of Twitter Users},
doi = {10.18653/v1/P17-1068}
}

@article{10.2307/2782461,
 ISSN = {00029602, 15375390},
 URL = {http://www.jstor.org/stable/2782461},
 abstract = {Many observers have asserted with little evidence that Americans' social opinions have become polarized. Using General Social Survey and National Election Survey social attitude items that have been repeated regularly over 20 years, the authors ask (1) Have Americans' opinions become more dispersed (higher variance)? (2) Have distributions become flatter or more bimodal (declining kurtosis)? (3) Have opinions become more ideologically constrained within and across opinion domains? (4) Have paired social groups become more different in their opinions? The authors find little evidence of polarization over the past two decades, with attitudes toward abortion and opinion differences between Republican and Democratic party identifiers the exceptional cases.},
 author = {Paul DiMaggio and John Evans and Bethany Bryson},
 journal = {American Journal of Sociology},
 number = {3},
 pages = {690--755},
 publisher = {University of Chicago Press},
 title = {Have American's Social Attitudes Become More Polarized?},
 urldate = {2022-08-19},
 volume = {102},
 year = {1996}
}


@inproceedings{9a75dad1ad314cb1ae6901cf09bd27b1,
title = "Political polarization in social media: Analysis of the 'Twitter political field' in Japan",
keywords = "Twitter, echo chambers, network analysis, political polarization, public sphere, topic modeling",
author = "Hiroki Takikawa and Kikuko Nagayoshi",
note = "Funding Information: This work was supported by JSPS KAKENHI Grant Numbers 16K13406, 16K04027 and 16K13347. We thank Yoshimichi Sato, Hiroshi Hamada, and Takafumi Ito for helpful comments and advice on an earlier version of this paper. Publisher Copyright: {\textcopyright} 2017 IEEE.; 5th IEEE International Conference on Big Data, Big Data 2017 ; Conference date: 11-12-2017 Through 14-12-2017",
year = "2017",
month = jul,
day = "1",
doi = "10.1109/BigData.2017.8258291",
language = "English",
series = "Proceedings - 2017 IEEE International Conference on Big Data, Big Data 2017",
publisher = "Institute of Electrical and Electronics Engineers Inc.",
pages = "3143--3150",
editor = "Jian-Yun Nie and Zoran Obradovic and Toyotaro Suzumura and Rumi Ghosh and Raghunath Nambiar and Chonggang Wang and Hui Zang and Ricardo Baeza-Yates and Ricardo Baeza-Yates and Xiaohua Hu and Jeremy Kepner and Alfredo Cuzzocrea and Jian Tang and Masashi Toyoda",
booktitle = "Proceedings - 2017 IEEE International Conference on Big Data, Big Data 2017",
}

@INPROCEEDINGS{9044124,
  author={Belcastro, Loris and Cantini, Riccardo and Marozzo, Fabrizio and Talia, Domenico and Trunfio, Paolo},
  booktitle={2019 15th International Conference on Semantics, Knowledge and Grids (SKG)}, 
  title={Discovering Political Polarization on Social Media: A Case Study}, 
  year={2019},
  volume={},
  number={},
  pages={182-189},
  doi={10.1109/SKG49510.2019.00038}}
  
@inproceedings{dash2022divided,
author = {Dash, Saloni and Mishra, Dibyendu and Shekhawat, Gazal and Pal, Joyojeet},
title = {Divided We Rule: Influencer Polarization on Twitter During Political Crises in India},
booktitle = {Proceedings of the International AAAI Conference on Web and Social Media},
year = {2022},
month = {June},
url = {https://www.microsoft.com/en-us/research/publication/divided-we-rule-influencer-polarization-on-twitter-during-political-crises-in-india/},
pages = {135-146},
}

@article{machado2018news,
  title={News and political information consumption in Brazil: Mapping the first round of the 2018 Brazilian presidential election on Twitter},
  author={Machado, Caio and Kira, Beatriz and Hirsch, Gustavo and Marchal, Nahema and Kollanyi, Bence and Howard, Philip N and Lederer, Thomas and Barash, Vlad},
  journal={The computational propaganda project. Algorithms, automation and digital politics. https://comprop. oii. ox. ac. uk/research/brazil2018},
  year={2018}
}

@article{Himelboim2013BirdsOA,
  title={Birds of a Feather Tweet Together: Integrating Network and Content Analyses to Examine Cross-Ideology Exposure on Twitter},
  author={Itai Himelboim and Stephen McCreery and Marc A. Smith},
  journal={J. Comput. Mediat. Commun.},
  year={2013},
  volume={18},
  pages={40-60}
}

@article{barbera_2015, title={Birds of the Same Feather Tweet Together: Bayesian Ideal Point Estimation Using Twitter Data}, volume={23}, DOI={10.1093/pan/mpu011}, number={1}, journal={Political Analysis}, publisher={Cambridge University Press}, author={Barberá, Pablo}, year={2015}, pages={76–91}}

@article{10.2307/45094371,
 ISSN = {0162895X, 14679221},
 URL = {http://www.jstor.org/stable/45094371},
 author = {Andrei Boutyline and Robb Willer},
 journal = {Political Psychology},
 number = {3},
 pages = {551--569},
 publisher = {[International Society of Political Psychology, Wiley]},
 title = {The Social Structure of Political Echo Chambers: Variation in Ideological Homophily in Online Networks},
 urldate = {2022-08-19},
 volume = {38},
 year = {2017}
}


@article{Colleoni,
author = {Colleoni, Elanor and Rozza, Alessandro and Arvidsson, Adam},
year = {2014},
month = {03},
pages = {},
title = {Echo Chamber or Public Sphere? Predicting Political Orientation and Measuring Political Homophily in Twitter Using Big Data},
volume = {64},
journal = {Journal of Communication},
doi = {10.1111/jcom.12084}
}
@article{HALBERSTAM201673,
title = {Homophily, group size, and the diffusion of political information in social networks: Evidence from Twitter},
journal = {Journal of Public Economics},
volume = {143},
pages = {73-88},
year = {2016},
issn = {0047-2727},
doi = {https://doi.org/10.1016/j.jpubeco.2016.08.011},
url = {https://www.sciencedirect.com/science/article/pii/S0047272716301001},
author = {Yosh Halberstam and Brian Knight},
keywords = {Political knowledge, Homophily, Social media, Ideological segregation},
}

@article{CHEN2021102348,
title = {Polarization of climate politics results from partisan sorting: Evidence from Finnish Twittersphere},
journal = {Global Environmental Change},
volume = {71},
pages = {102348},
year = {2021},
issn = {0959-3780},
doi = {https://doi.org/10.1016/j.gloenvcha.2021.102348},
url = {https://www.sciencedirect.com/science/article/pii/S0959378021001278},
author = {Ted Hsuan Yun Chen and Ali Salloum and Antti Gronow and Tuomas Ylä-Anttila and Mikko Kivelä},
keywords = {Climate politics, Political polarization, Partisan sorting, Issue alignment, Social networks},
abstract = {Prior research shows that public opinion on climate politics sorts along partisan lines. However, they leave open the question of whether climate politics and other politically salient issues exhibit tendencies for issue alignment, which the political polarization literature identifies as among the most deleterious aspects of polarization. Using a network approach and social media data from the Twitter platform, we study polarization of public opinion toward climate politics and ten other politically salient topics during the 2019 Finnish elections as the emergence of opposing groups in a public forum. We find that while climate politics is not particularly polarized compared to the other topics, it is subject to partisan sorting and issue alignment within the universalist-communitarian dimension of European politics that arose following the growth of right-wing populism. Notably, climate politics is consistently aligned with the immigration issue, and temporal trends indicate that this phenomenon will likely persist.}
}

@inproceedings{10.1145/3178876.3186130,
author = {Gillani, Nabeel and Yuan, Ann and Saveski, Martin and Vosoughi, Soroush and Roy, Deb},
title = {Me, My Echo Chamber, and I: Introspection on Social Media Polarization},
year = {2018},
isbn = {9781450356398},
publisher = {International World Wide Web Conferences Steering Committee},
address = {Republic and Canton of Geneva, CHE},
url = {https://doi.org/10.1145/3178876.3186130},
doi = {10.1145/3178876.3186130},
abstract = {Homophily - our tendency to surround ourselves with others who share our perspectives and opinions about the world - is both a part of human nature and an organizing principle underpinning many of our digital social networks. However, when it comes to politics or culture, homophily can amplify tribal mindsets and produce "echo chambers" that degrade the quality, safety, and diversity of discourse online. While several studies have empirically proven this point, few have explored how making users aware of the extent and nature of their political echo chambers influences their subsequent beliefs and actions. In this paper, we introduce Social Mirror, a social network visualization tool that enables a sample of Twitter users to explore the politically-active parts of their social network. We use Social Mirror to recruit Twitter users with a prior history of political discourse to a randomized experiment where we evaluate the effects of different treatments on participants' i) beliefs about their network connections, ii) the political diversity of who they choose to follow, and iii) the political alignment of the URLs they choose to share. While we see no effects on average political alignment of shared URLs, we find that recommending accounts of the opposite political ideology to follow reduces participants» beliefs in the political homogeneity of their network connections but still enhances their connection diversity one week after treatment. Conversely, participants who enhance their belief in the political homogeneity of their Twitter connections have less diverse network connections 2-3 weeks after treatment. We explore the implications of these disconnects between beliefs and actions on future efforts to promote healthier exchanges in our digital public spheres.},
booktitle = {Proceedings of the 2018 World Wide Web Conference},
pages = {823–831},
numpages = {9},
keywords = {randomized experiment, political polarization, social networks},
location = {Lyon, France},
series = {WWW '18}
}

@misc{https://doi.org/10.48550/arxiv.1409.8152,
  doi = {10.48550/ARXIV.1409.8152},
  
  url = {https://arxiv.org/abs/1409.8152},
  
  author = {Mejova, Yelena and Zhang, Amy X. and Diakopoulos, Nicholas and Castillo, Carlos},
  
  keywords = {Computers and Society (cs.CY), Computation and Language (cs.CL), FOS: Computer and information sciences, FOS: Computer and information sciences},
  
  title = {Controversy and Sentiment in Online News},
  
  publisher = {arXiv},
  
  year = {2014},
  
  copyright = {Creative Commons Attribution 3.0 Unported}
}


@INPROCEEDINGS{6785722,
  author={Weber, Ingmar and Garimella, Venkata R. Kiran and Batayneh, Alaa},
  booktitle={2013 IEEE/ACM International Conference on Advances in Social Networks Analysis and Mining (ASONAM 2013)}, 
  title={Secular vs. Islamist polarization in Egypt on Twitter}, 
  year={2013},
  volume={},
  number={},
  pages={290-297},
  doi={10.1109/ASONAM.2013.6785722}}
  
  @INPROCEEDINGS{8181508,
  author={Yang, Muheng and Wen, Xidao and Lin, Yu-Ru and Deng, Lingjia},
  booktitle={2017 IEEE 3rd International Conference on Collaboration and Internet Computing (CIC)}, 
  title={Quantifying Content Polarization on Twitter}, 
  year={2017},
  volume={},
  number={},
  pages={299-308},
  doi={10.1109/CIC.2017.00047}}
  
  
  @INPROCEEDINGS{8181508,
  author={Yang, Muheng and Wen, Xidao and Lin, Yu-Ru and Deng, Lingjia},
  booktitle={2017 IEEE 3rd International Conference on Collaboration and Internet Computing (CIC)}, 
  title={Quantifying Content Polarization on Twitter}, 
  year={2017},
  volume={},
  number={},
  pages={299-308},
  doi={10.1109/CIC.2017.00047}}
  
  
  @article{10.5555/944919.944937,
author = {Blei, David M. and Ng, Andrew Y. and Jordan, Michael I.},
title = {Latent Dirichlet Allocation},
year = {2003},
issue_date = {3/1/2003},
publisher = {JMLR.org},
volume = {3},
number = {null},
issn = {1532-4435},
abstract = {We describe latent Dirichlet allocation (LDA), a generative probabilistic model for collections of discrete data such as text corpora. LDA is a three-level hierarchical Bayesian model, in which each item of a collection is modeled as a finite mixture over an underlying set of topics. Each topic is, in turn, modeled as an infinite mixture over an underlying set of topic probabilities. In the context of text modeling, the topic probabilities provide an explicit representation of a document. We present efficient approximate inference techniques based on variational methods and an EM algorithm for empirical Bayes parameter estimation. We report results in document modeling, text classification, and collaborative filtering, comparing to a mixture of unigrams model and the probabilistic LSI model.},
journal = {J. Mach. Learn. Res.},
month = {mar},
pages = {993–1022},
numpages = {30}
}







@article{doi:10.1137/16M1087175,
author = {Fosdick, Bailey K. and Larremore, Daniel B. and Nishimura, Joel and Ugander, Johan},
title = {Configuring Random Graph Models with Fixed Degree Sequences},
journal = {SIAM Review},
volume = {60},
number = {2},
pages = {315-355},
year = {2018},
doi = {10.1137/16M1087175},

URL = { 
        https://doi.org/10.1137/16M1087175
    
},
eprint = { 
        https://doi.org/10.1137/16M1087175
    
}
,
  
}




@inproceedings{reimers2019sentencebert,
  added-at = {2020-10-27T13:25:14.000+0100},
  author = {Reimers, Nils and Gurevych, Iryna},
  biburl = {https://www.bibsonomy.org/bibtex/292afb49328e6651f17dc1e1b8c304aae/nosebrain},
  booktitle = {EMNLP/IJCNLP (1)},
  crossref = {conf/emnlp/2019-1},
  editor = {Inui, Kentaro and Jiang, Jing and Ng, Vincent and Wan, Xiaojun},
  ee = {https://doi.org/10.18653/v1/D19-1410},
  interhash = {762e8dacdc867460a7b200c6b4cd1b5c},
  intrahash = {92afb49328e6651f17dc1e1b8c304aae},
  isbn = {978-1-950737-90-1},
  keywords = {bert representation sentence sequence similarity},
  pages = {3980-3990},
  publisher = {Association for Computational Linguistics},
  timestamp = {2020-10-27T13:25:14.000+0100},
  title = {Sentence-BERT: Sentence Embeddings using Siamese BERT-Networks},
  url = {http://dblp.uni-trier.de/db/conf/emnlp/emnlp2019-1.html#ReimersG19},
  year = 2019
}



@inproceedings{shibayama-shinnou-2021-construction,
    title = "Construction and Evaluation of {J}apanese Sentence-{BERT} Models",
    author = "Shibayama, Naoki  and
      Shinnou, Hiroyuki",
    booktitle = "Proceedings of the 35th Pacific Asia Conference on Language, Information and Computation",
    month = "11",
    year = "2021",
    address = "Shanghai, China",
    publisher = "Association for Computational Lingustics",
    url = "https://aclanthology.org/2021.paclic-1.77",
    pages = "731--738",
}

@article{ctx31826023050006531,
address = {Washington, D.C. :},
issn = {0190-2725},
journal = {Social psychology quarterly},
lccn = {sn 99023429},
number = {2},
author = {Krackhardt, David},
publisher = {American Sociological Association,},
title = {Informal Networks and Organizational Crises: An Experimental Simulation},
volume = {51},
year = {1988},
}



@InProceedings{pmlr-v32-tang14,
  title = 	 {Understanding the Limiting Factors of Topic Modeling via Posterior Contraction Analysis},
  author = 	 {Tang, Jian and Meng, Zhaoshi and Nguyen, Xuanlong and Mei, Qiaozhu and Zhang, Ming},
  booktitle = 	 {Proceedings of the 31st International Conference on Machine Learning},
  pages = 	 {190--198},
  year = 	 {2014},
  editor = 	 {Xing, Eric P. and Jebara, Tony},
  volume = 	 {32},
  number =       {1},
  series = 	 {Proceedings of Machine Learning Research},
  address = 	 {Bejing, China},
  month = 	 {22--24 Jun},
  publisher =    {PMLR},
  pdf = 	 {http://proceedings.mlr.press/v32/tang14.pdf},
  url = 	 {https://proceedings.mlr.press/v32/tang14.html},
  abstract = 	 {Topic models such as the latent Dirichlet allocation (LDA) have become a standard staple in the modeling toolbox of machine learning. They have been applied to a vast variety of data sets, contexts, and tasks to varying degrees of success. However, to date there is almost no formal theory explicating the LDA’s behavior, and despite its familiarity there is very little systematic analysis of and guidance on the properties of the data that affect the inferential performance of the model. This paper seeks to address this gap, by providing a systematic analysis of factors which characterize the LDA’s performance.  We present theorems elucidating the posterior contraction rates of the topics as the amount of data increases, and a thorough supporting empirical study using synthetic and real data sets, including news and web-based articles and tweet messages. Based on these results we provide practical guidance on how to identify suitable data sets for topic models, and how to specify particular model parameters.}
}

@inproceedings{kudo-etal-2004-applying,
    title = "Applying Conditional Random Fields to {J}apanese Morphological Analysis",
    author = "Kudo, Taku  and
      Yamamoto, Kaoru  and
      Matsumoto, Yuji",
    booktitle = "Proceedings of the 2004 Conference on Empirical Methods in Natural Language Processing",
    month = jul,
    year = "2004",
    address = "Barcelona, Spain",
    publisher = "Association for Computational Linguistics",
    url = "https://aclanthology.org/W04-3230",
    pages = "230--237",
}

@article{uchida,
author = {Uchida, Kazuki and Toriumi, Fujio and Sakaki, Takeshi},
year = {2019},
month = {11},
pages = {1-14},
title = {Comparative evaluation of two approaches for retweet clustering: A text-based method and graph-based method},
volume = {17},
journal = {Web Intelligence},
doi = {10.3233/WEB-190418}
}

@inproceedings{preotiuc-pietro-etal-2017-beyond,
    title = "Beyond Binary Labels: Political Ideology Prediction of {T}witter Users",
    author = "Preo{\c{t}}iuc-Pietro, Daniel  and
      Liu, Ye  and
      Hopkins, Daniel  and
      Ungar, Lyle",
    booktitle = "Proceedings of the 55th Annual Meeting of the Association for Computational Linguistics (Volume 1: Long Papers)",
    month = jul,
    year = "2017",
    address = "Vancouver, Canada",
    publisher = "Association for Computational Linguistics",
    url = "https://aclanthology.org/P17-1068",
    doi = "10.18653/v1/P17-1068",
    pages = "729--740",
    abstract = "Automatic political orientation prediction from social media posts has to date proven successful only in distinguishing between publicly declared liberals and conservatives in the US. This study examines users{'} political ideology using a seven-point scale which enables us to identify politically moderate and neutral users {--} groups which are of particular interest to political scientists and pollsters. Using a novel data set with political ideology labels self-reported through surveys, our goal is two-fold: a) to characterize the groups of politically engaged users through language use on Twitter; b) to build a fine-grained model that predicts political ideology of unseen users. Our results identify differences in both political leaning and engagement and the extent to which each group tweets using political keywords. Finally, we demonstrate how to improve ideology prediction accuracy by exploiting the relationships between the user groups.",
}

@inproceedings{10.1145/3394486.3403275,
author = {Xiao, Zhiping and Song, Weiping and Xu, Haoyan and Ren, Zhicheng and Sun, Yizhou},
title = {TIMME: Twitter Ideology-Detection via Multi-Task Multi-Relational Embedding},
year = {2020},
isbn = {9781450379984},
publisher = {Association for Computing Machinery},
address = {New York, NY, USA},
url = {https://doi.org/10.1145/3394486.3403275},
doi = {10.1145/3394486.3403275},
abstract = {We aim at solving the problem of predicting people's ideology, or political tendency. We estimate it by using Twitter data, and formalize it as a classification problem. Ideology-detection has long been a challenging yet important problem. Certain groups, such as the policy makers, rely on it to make wise decisions. Back in the old days when labor-intensive survey-studies were needed to collect public opinions, analyzing ordinary citizens' political tendencies was uneasy. The rise of social medias, such as Twitter, has enabled us to gather ordinary citizen's data easily. However, the incompleteness of the labels and the features in social network datasets is tricky, not to mention the enormous data size and the heterogeneousity. The data differ dramatically from many commonly-used datasets, thus brings unique challenges. In our work, first we built our own datasets from Twitter. Next, we proposed TIMME, a multi-task multi-relational embedding model, that works efficiently on sparsely-labeled heterogeneous real-world dataset. It could also handle the incompleteness of the input features. Experimental results showed that TIMME is overall better than the state-of-the-art models for ideology detection on Twitter. Our findings include: links can lead to good classification outcomes without text; conservative voice is under-represented on Twitter; follow is the most important relation to predict ideology; retweet and mention enhance a higher chance of like, etc. Last but not least, TIMME could be extended to other datasets and tasks in theory.},
booktitle = {Proceedings of the 26th ACM SIGKDD International Conference on Knowledge Discovery \& Data Mining},
pages = {2258–2268},
numpages = {11},
keywords = {social network analysis, ideology detection, graph convolutional networks, heterogeneous information network, multi-task learning},
location = {Virtual Event, CA, USA},
series = {KDD '20}
}

@inproceedings{baly-etal-2019-multi,
    title = "Multi-Task Ordinal Regression for Jointly Predicting the Trustworthiness and the Leading Political Ideology of News Media",
    author = "Baly, Ramy  and
      Karadzhov, Georgi  and
      Saleh, Abdelrhman  and
      Glass, James  and
      Nakov, Preslav",
    booktitle = "Proceedings of the 2019 Conference of the North {A}merican Chapter of the Association for Computational Linguistics: Human Language Technologies, Volume 1 (Long and Short Papers)",
    month = jun,
    year = "2019",
    address = "Minneapolis, Minnesota",
    publisher = "Association for Computational Linguistics",
    url = "https://aclanthology.org/N19-1216",
    doi = "10.18653/v1/N19-1216",
    pages = "2109--2116",
    abstract = "In the context of fake news, bias, and propaganda, we study two important but relatively under-explored problems: (i) trustworthiness estimation (on a 3-point scale) and (ii) political ideology detection (left/right bias on a 7-point scale) of entire news outlets, as opposed to evaluating individual articles. In particular, we propose a multi-task ordinal regression framework that models the two problems jointly. This is motivated by the observation that hyper-partisanship is often linked to low trustworthiness, e.g., appealing to emotions rather than sticking to the facts, while center media tend to be generally more impartial and trustworthy. We further use several auxiliary tasks, modeling centrality, hyper-partisanship, as well as left-vs.-right bias on a coarse-grained scale. The evaluation results show sizable performance gains by the joint models over models that target the problems in isolation.",
}

@article{Rizoiu_Graham_Zhang_Zhang_Ackland_Xie_2018, title={#DebateNight: The Role and Influence of Socialbots on Twitter During the 1st 2016 U.S. Presidential Debate}, volume={12}, url={https://ojs.aaai.org/index.php/ICWSM/article/view/15029}, abstractNote={ &lt;p&gt; Serious concerns have been raised about the role of `socialbots’ in manipulating public opinion and influencing the outcome of elections by retweeting partisan content to increase its reach. Here we analyze the role and influence of socialbots on Twitter by determining how they contribute to retweet diffusions. We collect a large dataset of tweets during the 1st U.S. presidential debate in 2016 and we analyze its 1.5 million users from three perspectives: user influence, political behavior (partisanship and engagement) and botness. First, we define a measure of user influence based on the user’s active contributions to information diffusions, i.e. their tweets and retweets. Given that Twitter does not expose the retweet structure -- it associates all retweets with the original tweet -- we model the latent diffusion structure using only tweet time and user features, and we implement a scalable novel approach to estimate influence over all possible unfoldings. Next, we use partisan hashtag analysis to quantify user political polarization and engagement. Finally, we use the BotOrNot API to measure user botness (the likelihood of being a bot). We build a two-dimensional &quot;polarization map&quot; that allows for a nuanced analysis of the interplay between botness, partisanship and influence. We find that not only are socialbots more active on Twitter -- starting more retweet cascades and retweeting more -- but they are 2.5 times more influential than humans, and more politically engaged. Moreover, pro-Republican bots are both more influential and more politically engaged than their pro-Democrat counterparts. However we caution against blanket statements that software designed to appear human dominates politics-related activity on Twitter. Firstly, it is known that accounts controlled by teams of humans (e.g. organizational accounts) are often identified as bots. Secondly, we find that many highly influential Twitter users are in fact pro-Democrat and that most pro-Republican users are mid-influential and likely to be human (low botness). &lt;/p&gt; }, number={1}, journal={Proceedings of the International AAAI Conference on Web and Social Media}, author={Rizoiu, Marian-Andrei and Graham, Timothy and Zhang, Rui and Zhang, Yifei and Ackland, Robert and Xie, Lexing}, year={2018}, month={Jun.} }


@Article{rf,
title = {Classification and Regression by randomForest},
author = {Andy Liaw and Matthew Wiener},
journal = {R News},
year = {2002},
volume = {2},
number = {3},
pages = {18-22},
url = {https://CRAN.R-project.org/doc/Rnews/},
}

@inproceedings{10.1145/1159913.1159930,
author = {Mahadevan, Priya and Krioukov, Dmitri and Fall, Kevin and Vahdat, Amin},
title = {Systematic Topology Analysis and Generation Using Degree Correlations},
year = {2006},
isbn = {1595933085},
publisher = {Association for Computing Machinery},
address = {New York, NY, USA},
url = {https://doi.org/10.1145/1159913.1159930},
doi = {10.1145/1159913.1159930},
booktitle = {Proceedings of the 2006 Conference on Applications, Technologies, Architectures, and Protocols for Computer Communications},
pages = {135–146},
numpages = {12},
keywords = {degree correlations, network topology},
location = {Pisa, Italy},
series = {SIGCOMM '06}

@article{10.1145/3209581,
author = {Baeza-Yates, Ricardo},
title = {Bias on the Web},
year = {2018},
issue_date = {June 2018},
publisher = {Association for Computing Machinery},
address = {New York, NY, USA},
volume = {61},
number = {6},
issn = {0001-0782},
url = {https://doi.org/10.1145/3209581},
doi = {10.1145/3209581},
abstract = {Bias in Web data and use taints the algorithms behind Web-based applications, delivering equally biased results.},
journal = {Communications of the ACM},
month = may,
pages = {54–61},
numpages = {8}
}

@article{fa2,
author = {Jacomy, Mathieu and Venturini, Tommaso and Heymann, Sebastien and Bastian, Mathieu},
year = {2014},
month = {06},
pages = {e98679},
title = {ForceAtlas2, a Continuous Graph Layout Algorithm for Handy Network Visualization Designed for the Gephi Software},
volume = {9},
journal = {PloS one},
doi = {10.1371/journal.pone.0098679}
}

@article{Nyhan2010,
  abstract = {An extensive literature addresses citizen ignorance, but very little research focuses on misperceptions. Can these false or unsubstantiated beliefs about politics be corrected? Previous studies have not tested the efficacy of corrections in a realistic format. We conducted four experiments in which subjects read mock news articles that included either a misleading claim from a politician, or a misleading claim and a correction. Results indicate that corrections frequently fail to reduce misperceptions among the targeted ideological group. We also document several instances of a ``backfire effect'' in which corrections actually increase misperceptions among the group in question.},
  added-at = {2018-04-03T23:53:55.000+0200},
  author = {Nyhan, Brendan and Reifler, Jason},
  biburl = {https://www.bibsonomy.org/bibtex/2fb6a36e74c1d827ff8e7abb91699ef2d/psychrec},
  day = 01,
  description = {When Corrections Fail: The Persistence of Political Misperceptions | SpringerLink},
  doi = {10.1007/s11109-010-9112-2},
  interhash = {f1e68aafbf02cca80f2b425b02bd262e},
  intrahash = {fb6a36e74c1d827ff8e7abb91699ef2d},
  issn = {1573-6687},
  journal = {Political Behavior},
  keywords = {backfire-effect psychology},
  month = jun,
  number = 2,
  pages = {303--330},
  timestamp = {2018-04-03T23:53:55.000+0200},
  title = {When Corrections Fail: The Persistence of Political Misperceptions},
  url = {https://doi.org/10.1007/s11109-010-9112-2},
  volume = 32,
  year = 2010
}




\begin{thebibliography}{40}


\ifx \showCODEN    \undefined \def \showCODEN     #1{\unskip}     \fi
\ifx \showDOI      \undefined \def \showDOI       #1{#1}\fi
\ifx \showISBNx    \undefined \def \showISBNx     #1{\unskip}     \fi
\ifx \showISBNxiii \undefined \def \showISBNxiii  #1{\unskip}     \fi
\ifx \showISSN     \undefined \def \showISSN      #1{\unskip}     \fi
\ifx \showLCCN     \undefined \def \showLCCN      #1{\unskip}     \fi
\ifx \shownote     \undefined \def \shownote      #1{#1}          \fi
\ifx \showarticletitle \undefined \def \showarticletitle #1{#1}   \fi
\ifx \showURL      \undefined \def \showURL       {\relax}        \fi
\providecommand\bibfield[2]{#2}
\providecommand\bibinfo[2]{#2}
\providecommand\natexlab[1]{#1}
\providecommand\showeprint[2][]{arXiv:#2}

\bibitem[\protect\citeauthoryear{Baeza-Yates}{Baeza-Yates}{2018}]%
        {10.1145/3209581}
\bibfield{author}{\bibinfo{person}{Ricardo Baeza-Yates}.}
  \bibinfo{year}{2018}\natexlab{}.
\newblock \showarticletitle{Bias on the Web}.
\newblock \bibinfo{journal}{\emph{Commun. ACM}} \bibinfo{volume}{61},
  \bibinfo{number}{6} (\bibinfo{date}{May} \bibinfo{year}{2018}),
  \bibinfo{pages}{54–61}.
\newblock
\showISSN{0001-0782}
\urldef\tempurl%
\url{https://doi.org/10.1145/3209581}
\showDOI{\tempurl}


\bibitem[\protect\citeauthoryear{Baly, Karadzhov, Saleh, Glass, and Nakov}{Baly
  et~al\mbox{.}}{2019}]%
        {baly-etal-2019-multi}
\bibfield{author}{\bibinfo{person}{Ramy Baly}, \bibinfo{person}{Georgi
  Karadzhov}, \bibinfo{person}{Abdelrhman Saleh}, \bibinfo{person}{James
  Glass}, {and} \bibinfo{person}{Preslav Nakov}.}
  \bibinfo{year}{2019}\natexlab{}.
\newblock \showarticletitle{Multi-Task Ordinal Regression for Jointly
  Predicting the Trustworthiness and the Leading Political Ideology of News
  Media}. In \bibinfo{booktitle}{\emph{Proceedings of the 2019 Conference of
  the North {A}merican Chapter of the Association for Computational
  Linguistics: Human Language Technologies, Volume 1 (Long and Short Papers)}}.
  \bibinfo{publisher}{Association for Computational Linguistics},
  \bibinfo{address}{Minneapolis, Minnesota}, \bibinfo{pages}{2109--2116}.
\newblock
\urldef\tempurl%
\url{https://doi.org/10.18653/v1/N19-1216}
\showDOI{\tempurl}


\bibitem[\protect\citeauthoryear{Barberá}{Barberá}{2015}]%
        {barbera_2015}
\bibfield{author}{\bibinfo{person}{Pablo Barberá}.}
  \bibinfo{year}{2015}\natexlab{}.
\newblock \showarticletitle{Birds of the Same Feather Tweet Together: Bayesian
  Ideal Point Estimation Using Twitter Data}.
\newblock \bibinfo{journal}{\emph{Political Analysis}} \bibinfo{volume}{23},
  \bibinfo{number}{1} (\bibinfo{year}{2015}), \bibinfo{pages}{76–91}.
\newblock
\urldef\tempurl%
\url{https://doi.org/10.1093/pan/mpu011}
\showDOI{\tempurl}


\bibitem[\protect\citeauthoryear{Barberá, Jost, Nagler, Tucker, and
  Bonneau}{Barberá et~al\mbox{.}}{2015}]%
        {Barbera}
\bibfield{author}{\bibinfo{person}{Pablo Barberá}, \bibinfo{person}{John
  Jost}, \bibinfo{person}{Jonathan Nagler}, \bibinfo{person}{Joshua Tucker},
  {and} \bibinfo{person}{Richard Bonneau}.} \bibinfo{year}{2015}\natexlab{}.
\newblock \showarticletitle{Tweeting From Left to Right: Is Online Political
  Communication More Than an Echo Chamber?}
\newblock \bibinfo{journal}{\emph{Psychological science}}  \bibinfo{volume}{26}
  (\bibinfo{date}{08} \bibinfo{year}{2015}).
\newblock
\urldef\tempurl%
\url{https://doi.org/10.1177/0956797615594620}
\showDOI{\tempurl}


\bibitem[\protect\citeauthoryear{Blei, Ng, and Jordan}{Blei
  et~al\mbox{.}}{2003}]%
        {10.5555/944919.944937}
\bibfield{author}{\bibinfo{person}{David~M. Blei}, \bibinfo{person}{Andrew~Y.
  Ng}, {and} \bibinfo{person}{Michael~I. Jordan}.}
  \bibinfo{year}{2003}\natexlab{}.
\newblock \showarticletitle{Latent Dirichlet Allocation}.
\newblock \bibinfo{journal}{\emph{J. Mach. Learn. Res.}} \bibinfo{volume}{3},
  \bibinfo{number}{null} (\bibinfo{date}{mar} \bibinfo{year}{2003}),
  \bibinfo{pages}{993–1022}.
\newblock
\showISSN{1532-4435}


\bibitem[\protect\citeauthoryear{Boutyline and Willer}{Boutyline and
  Willer}{2017}]%
        {10.2307/45094371}
\bibfield{author}{\bibinfo{person}{Andrei Boutyline} {and}
  \bibinfo{person}{Robb Willer}.} \bibinfo{year}{2017}\natexlab{}.
\newblock \showarticletitle{The Social Structure of Political Echo Chambers:
  Variation in Ideological Homophily in Online Networks}.
\newblock \bibinfo{journal}{\emph{Political Psychology}} \bibinfo{volume}{38},
  \bibinfo{number}{3} (\bibinfo{year}{2017}), \bibinfo{pages}{551--569}.
\newblock
\showISSN{0162895X, 14679221}
\urldef\tempurl%
\url{http://www.jstor.org/stable/45094371}
\showURL{%
\tempurl}


\bibitem[\protect\citeauthoryear{Chen, Salloum, Gronow, Ylä-Anttila, and
  Kivelä}{Chen et~al\mbox{.}}{2021}]%
        {CHEN2021102348}
\bibfield{author}{\bibinfo{person}{Ted Hsuan~Yun Chen}, \bibinfo{person}{Ali
  Salloum}, \bibinfo{person}{Antti Gronow}, \bibinfo{person}{Tuomas
  Ylä-Anttila}, {and} \bibinfo{person}{Mikko Kivelä}.}
  \bibinfo{year}{2021}\natexlab{}.
\newblock \showarticletitle{Polarization of climate politics results from
  partisan sorting: Evidence from Finnish Twittersphere}.
\newblock \bibinfo{journal}{\emph{Global Environmental Change}}
  \bibinfo{volume}{71} (\bibinfo{year}{2021}), \bibinfo{pages}{102348}.
\newblock
\showISSN{0959-3780}
\urldef\tempurl%
\url{https://doi.org/10.1016/j.gloenvcha.2021.102348}
\showDOI{\tempurl}


\bibitem[\protect\citeauthoryear{Colleoni, Rozza, and Arvidsson}{Colleoni
  et~al\mbox{.}}{2014}]%
        {Colleoni}
\bibfield{author}{\bibinfo{person}{Elanor Colleoni},
  \bibinfo{person}{Alessandro Rozza}, {and} \bibinfo{person}{Adam Arvidsson}.}
  \bibinfo{year}{2014}\natexlab{}.
\newblock \showarticletitle{Echo Chamber or Public Sphere? Predicting Political
  Orientation and Measuring Political Homophily in Twitter Using Big Data}.
\newblock \bibinfo{journal}{\emph{Journal of Communication}}
  \bibinfo{volume}{64} (\bibinfo{date}{03} \bibinfo{year}{2014}).
\newblock
\urldef\tempurl%
\url{https://doi.org/10.1111/jcom.12084}
\showDOI{\tempurl}


\bibitem[\protect\citeauthoryear{Conover, Ratkiewicz, Francisco, Goncalves,
  Menczer, and Flammini}{Conover et~al\mbox{.}}{2021}]%
        {Conover_Ratkiewicz_Francisco_Goncalves_Menczer_Flammini_2021}
\bibfield{author}{\bibinfo{person}{Michael Conover}, \bibinfo{person}{Jacob
  Ratkiewicz}, \bibinfo{person}{Matthew Francisco}, \bibinfo{person}{Bruno
  Goncalves}, \bibinfo{person}{Filippo Menczer}, {and}
  \bibinfo{person}{Alessandro Flammini}.} \bibinfo{year}{2021}\natexlab{}.
\newblock \showarticletitle{Political Polarization on Twitter}.
\newblock \bibinfo{journal}{\emph{Proceedings of the International AAAI
  Conference on Web and Social Media}} \bibinfo{volume}{5}, \bibinfo{number}{1}
  (\bibinfo{date}{Aug.} \bibinfo{year}{2021}), \bibinfo{pages}{89--96}.
\newblock
\urldef\tempurl%
\url{https://ojs.aaai.org/index.php/ICWSM/article/view/14126}
\showURL{%
\tempurl}


\bibitem[\protect\citeauthoryear{Conover, Goncalves, Ratkiewicz, Flammini, and
  Menczer}{Conover et~al\mbox{.}}{2011}]%
        {6113114}
\bibfield{author}{\bibinfo{person}{Michael~D. Conover}, \bibinfo{person}{Bruno
  Goncalves}, \bibinfo{person}{Jacob Ratkiewicz}, \bibinfo{person}{Alessandro
  Flammini}, {and} \bibinfo{person}{Filippo Menczer}.}
  \bibinfo{year}{2011}\natexlab{}.
\newblock \showarticletitle{Predicting the Political Alignment of Twitter
  Users}. In \bibinfo{booktitle}{\emph{2011 IEEE Third International Conference
  on Privacy, Security, Risk and Trust and 2011 IEEE Third International
  Conference on Social Computing}}. \bibinfo{pages}{192--199}.
\newblock
\urldef\tempurl%
\url{https://doi.org/10.1109/PASSAT/SocialCom.2011.34}
\showDOI{\tempurl}


\bibitem[\protect\citeauthoryear{Del~Vicario, Quattrociocchi, Scala, and
  Zollo}{Del~Vicario et~al\mbox{.}}{2019}]%
        {polarfake}
\bibfield{author}{\bibinfo{person}{Michela Del~Vicario},
  \bibinfo{person}{Walter Quattrociocchi}, \bibinfo{person}{Antonio Scala},
  {and} \bibinfo{person}{Fabiana Zollo}.} \bibinfo{year}{2019}\natexlab{}.
\newblock \showarticletitle{Polarization and Fake News: Early Warning of
  Potential Misinformation Targets}.
\newblock \bibinfo{journal}{\emph{ACM Transactions on the Web}}
  \bibinfo{volume}{13} (\bibinfo{date}{03} \bibinfo{year}{2019}).
\newblock
\urldef\tempurl%
\url{https://doi.org/10.1145/3316809}
\showDOI{\tempurl}


\bibitem[\protect\citeauthoryear{DiMaggio, Evans, and Bryson}{DiMaggio
  et~al\mbox{.}}{1996}]%
        {10.2307/2782461}
\bibfield{author}{\bibinfo{person}{Paul DiMaggio}, \bibinfo{person}{John
  Evans}, {and} \bibinfo{person}{Bethany Bryson}.}
  \bibinfo{year}{1996}\natexlab{}.
\newblock \showarticletitle{Have American's Social Attitudes Become More
  Polarized?}
\newblock \bibinfo{journal}{\emph{Amer. J. Sociology}} \bibinfo{volume}{102},
  \bibinfo{number}{3} (\bibinfo{year}{1996}), \bibinfo{pages}{690--755}.
\newblock
\showISSN{00029602, 15375390}
\urldef\tempurl%
\url{http://www.jstor.org/stable/2782461}
\showURL{%
\tempurl}


\bibitem[\protect\citeauthoryear{Garimella, Morales, Gionis, and
  Mathioudakis}{Garimella et~al\mbox{.}}{2018}]%
        {10.1145/3140565}
\bibfield{author}{\bibinfo{person}{Kiran Garimella}, \bibinfo{person}{Gianmarco
  De~Francisci Morales}, \bibinfo{person}{Aristides Gionis}, {and}
  \bibinfo{person}{Michael Mathioudakis}.} \bibinfo{year}{2018}\natexlab{}.
\newblock \showarticletitle{Quantifying Controversy on Social Media}.
\newblock \bibinfo{journal}{\emph{Trans. Soc. Comput.}} \bibinfo{volume}{1},
  \bibinfo{number}{1}, Article \bibinfo{articleno}{3} (\bibinfo{date}{jan}
  \bibinfo{year}{2018}), \bibinfo{numpages}{27}~pages.
\newblock
\showISSN{2469-7818}
\urldef\tempurl%
\url{https://doi.org/10.1145/3140565}
\showDOI{\tempurl}


\bibitem[\protect\citeauthoryear{Garimella and Weber}{Garimella and
  Weber}{2017}]%
        {Garimella_Weber_2017}
\bibfield{author}{\bibinfo{person}{Venkata Rama~Kiran Garimella} {and}
  \bibinfo{person}{Ingmar Weber}.} \bibinfo{year}{2017}\natexlab{}.
\newblock \showarticletitle{A Long-Term Analysis of Polarization on Twitter}.
\newblock \bibinfo{journal}{\emph{Proceedings of the International AAAI
  Conference on Web and Social Media}} \bibinfo{volume}{11},
  \bibinfo{number}{1} (\bibinfo{date}{May} \bibinfo{year}{2017}),
  \bibinfo{pages}{528--531}.
\newblock
\urldef\tempurl%
\url{https://ojs.aaai.org/index.php/ICWSM/article/view/14918}
\showURL{%
\tempurl}


\bibitem[\protect\citeauthoryear{Gillani, Yuan, Saveski, Vosoughi, and
  Roy}{Gillani et~al\mbox{.}}{2018}]%
        {10.1145/3178876.3186130}
\bibfield{author}{\bibinfo{person}{Nabeel Gillani}, \bibinfo{person}{Ann Yuan},
  \bibinfo{person}{Martin Saveski}, \bibinfo{person}{Soroush Vosoughi}, {and}
  \bibinfo{person}{Deb Roy}.} \bibinfo{year}{2018}\natexlab{}.
\newblock \showarticletitle{Me, My Echo Chamber, and I: Introspection on Social
  Media Polarization}. In \bibinfo{booktitle}{\emph{Proceedings of the 2018
  World Wide Web Conference}} (Lyon, France) \emph{(\bibinfo{series}{WWW
  '18})}. \bibinfo{publisher}{International World Wide Web Conferences Steering
  Committee}, \bibinfo{address}{Republic and Canton of Geneva, CHE},
  \bibinfo{pages}{823–831}.
\newblock
\showISBNx{9781450356398}
\urldef\tempurl%
\url{https://doi.org/10.1145/3178876.3186130}
\showDOI{\tempurl}


\bibitem[\protect\citeauthoryear{Halberstam and Knight}{Halberstam and
  Knight}{2016}]%
        {HALBERSTAM201673}
\bibfield{author}{\bibinfo{person}{Yosh Halberstam} {and}
  \bibinfo{person}{Brian Knight}.} \bibinfo{year}{2016}\natexlab{}.
\newblock \showarticletitle{Homophily, group size, and the diffusion of
  political information in social networks: Evidence from Twitter}.
\newblock \bibinfo{journal}{\emph{Journal of Public Economics}}
  \bibinfo{volume}{143} (\bibinfo{year}{2016}), \bibinfo{pages}{73--88}.
\newblock
\showISSN{0047-2727}
\urldef\tempurl%
\url{https://doi.org/10.1016/j.jpubeco.2016.08.011}
\showDOI{\tempurl}


\bibitem[\protect\citeauthoryear{He, Mokhberian, Camara, Abeliuk, and
  Lerman}{He et~al\mbox{.}}{2021}]%
        {He2021DetectingPT}
\bibfield{author}{\bibinfo{person}{Zihao He}, \bibinfo{person}{Negar
  Mokhberian}, \bibinfo{person}{Antonio Camara}, \bibinfo{person}{Andr{\'e}s
  Abeliuk}, {and} \bibinfo{person}{Kristina Lerman}.}
  \bibinfo{year}{2021}\natexlab{}.
\newblock \showarticletitle{Detecting Polarized Topics Using Partisanship-aware
  Contextualized Topic Embeddings}. In \bibinfo{booktitle}{\emph{EMNLP}}.
\newblock


\bibitem[\protect\citeauthoryear{Himelboim, McCreery, and Smith}{Himelboim
  et~al\mbox{.}}{2013}]%
        {Himelboim2013BirdsOA}
\bibfield{author}{\bibinfo{person}{Itai Himelboim}, \bibinfo{person}{Stephen
  McCreery}, {and} \bibinfo{person}{Marc~A. Smith}.}
  \bibinfo{year}{2013}\natexlab{}.
\newblock \showarticletitle{Birds of a Feather Tweet Together: Integrating
  Network and Content Analyses to Examine Cross-Ideology Exposure on Twitter}.
\newblock \bibinfo{journal}{\emph{J. Comput. Mediat. Commun.}}
  \bibinfo{volume}{18} (\bibinfo{year}{2013}), \bibinfo{pages}{40--60}.
\newblock


\bibitem[\protect\citeauthoryear{Jacomy, Venturini, Heymann, and
  Bastian}{Jacomy et~al\mbox{.}}{2014}]%
        {fa2}
\bibfield{author}{\bibinfo{person}{Mathieu Jacomy}, \bibinfo{person}{Tommaso
  Venturini}, \bibinfo{person}{Sebastien Heymann}, {and}
  \bibinfo{person}{Mathieu Bastian}.} \bibinfo{year}{2014}\natexlab{}.
\newblock \showarticletitle{ForceAtlas2, a Continuous Graph Layout Algorithm
  for Handy Network Visualization Designed for the Gephi Software}.
\newblock \bibinfo{journal}{\emph{PloS one}}  \bibinfo{volume}{9}
  (\bibinfo{date}{06} \bibinfo{year}{2014}), \bibinfo{pages}{e98679}.
\newblock
\urldef\tempurl%
\url{https://doi.org/10.1371/journal.pone.0098679}
\showDOI{\tempurl}


\bibitem[\protect\citeauthoryear{Karypis and Kumar}{Karypis and Kumar}{1998}]%
        {doi:10.1137/S1064827595287997}
\bibfield{author}{\bibinfo{person}{George Karypis} {and} \bibinfo{person}{Vipin
  Kumar}.} \bibinfo{year}{1998}\natexlab{}.
\newblock \showarticletitle{A Fast and High Quality Multilevel Scheme for
  Partitioning Irregular Graphs}.
\newblock \bibinfo{journal}{\emph{SIAM Journal on Scientific Computing}}
  \bibinfo{volume}{20}, \bibinfo{number}{1} (\bibinfo{year}{1998}),
  \bibinfo{pages}{359--392}.
\newblock
\urldef\tempurl%
\url{https://doi.org/10.1137/S1064827595287997}
\showDOI{\tempurl}
\showeprint{https://doi.org/10.1137/S1064827595287997}


\bibitem[\protect\citeauthoryear{Krackhardt}{Krackhardt}{1988}]%
        {ctx31826023050006531}
\bibfield{author}{\bibinfo{person}{David Krackhardt}.}
  \bibinfo{year}{1988}\natexlab{}.
\newblock \showarticletitle{Informal Networks and Organizational Crises: An
  Experimental Simulation}.
\newblock \bibinfo{journal}{\emph{Social psychology quarterly}}
  \bibinfo{volume}{51}, \bibinfo{number}{2} (\bibinfo{year}{1988}).
\newblock
\showISSN{0190-2725}
\showLCCN{sn 99023429}


\bibitem[\protect\citeauthoryear{Krackhardt and Stern}{Krackhardt and
  Stern}{1988}]%
        {10.2307/2786835}
\bibfield{author}{\bibinfo{person}{David Krackhardt} {and}
  \bibinfo{person}{Robert~N. Stern}.} \bibinfo{year}{1988}\natexlab{}.
\newblock \showarticletitle{Informal Networks and Organizational Crises: An
  Experimental Simulation}.
\newblock \bibinfo{journal}{\emph{Social Psychology Quarterly}}
  \bibinfo{volume}{51}, \bibinfo{number}{2} (\bibinfo{year}{1988}),
  \bibinfo{pages}{123--140}.
\newblock
\showISSN{01902725}
\urldef\tempurl%
\url{http://www.jstor.org/stable/2786835}
\showURL{%
\tempurl}


\bibitem[\protect\citeauthoryear{Kudo, Yamamoto, and Matsumoto}{Kudo
  et~al\mbox{.}}{2004}]%
        {kudo-etal-2004-applying}
\bibfield{author}{\bibinfo{person}{Taku Kudo}, \bibinfo{person}{Kaoru
  Yamamoto}, {and} \bibinfo{person}{Yuji Matsumoto}.}
  \bibinfo{year}{2004}\natexlab{}.
\newblock \showarticletitle{Applying Conditional Random Fields to {J}apanese
  Morphological Analysis}. In \bibinfo{booktitle}{\emph{Proceedings of the 2004
  Conference on Empirical Methods in Natural Language Processing}}.
  \bibinfo{publisher}{Association for Computational Linguistics},
  \bibinfo{address}{Barcelona, Spain}, \bibinfo{pages}{230--237}.
\newblock
\urldef\tempurl%
\url{https://aclanthology.org/W04-3230}
\showURL{%
\tempurl}


\bibitem[\protect\citeauthoryear{Liaw and Wiener}{Liaw and Wiener}{2002}]%
        {rf}
\bibfield{author}{\bibinfo{person}{Andy Liaw} {and} \bibinfo{person}{Matthew
  Wiener}.} \bibinfo{year}{2002}\natexlab{}.
\newblock \showarticletitle{Classification and Regression by randomForest}.
\newblock \bibinfo{journal}{\emph{R News}} \bibinfo{volume}{2},
  \bibinfo{number}{3} (\bibinfo{year}{2002}), \bibinfo{pages}{18--22}.
\newblock
\urldef\tempurl%
\url{https://CRAN.R-project.org/doc/Rnews/}
\showURL{%
\tempurl}


\bibitem[\protect\citeauthoryear{Mahadevan, Krioukov, Fall, and
  Vahdat}{Mahadevan et~al\mbox{.}}{2006}]%
        {10.1145/1159913.1159930}
\bibfield{author}{\bibinfo{person}{Priya Mahadevan}, \bibinfo{person}{Dmitri
  Krioukov}, \bibinfo{person}{Kevin Fall}, {and} \bibinfo{person}{Amin
  Vahdat}.} \bibinfo{year}{2006}\natexlab{}.
\newblock \showarticletitle{Systematic Topology Analysis and Generation Using
  Degree Correlations}. In \bibinfo{booktitle}{\emph{Proceedings of the 2006
  Conference on Applications, Technologies, Architectures, and Protocols for
  Computer Communications}} (Pisa, Italy) \emph{(\bibinfo{series}{SIGCOMM
  '06})}. \bibinfo{publisher}{Association for Computing Machinery},
  \bibinfo{address}{New York, NY, USA}, \bibinfo{pages}{135–146}.
\newblock
\showISBNx{1595933085}
\urldef\tempurl%
\url{https://doi.org/10.1145/1159913.1159930}
\showDOI{\tempurl}


\bibitem[\protect\citeauthoryear{Mejova, Zhang, Diakopoulos, and
  Castillo}{Mejova et~al\mbox{.}}{2014a}]%
        {https://doi.org/10.48550/arxiv.1409.8152}
\bibfield{author}{\bibinfo{person}{Yelena Mejova}, \bibinfo{person}{Amy~X.
  Zhang}, \bibinfo{person}{Nicholas Diakopoulos}, {and} \bibinfo{person}{Carlos
  Castillo}.} \bibinfo{year}{2014}\natexlab{a}.
\newblock \bibinfo{title}{Controversy and Sentiment in Online News}.
\newblock
\newblock
\urldef\tempurl%
\url{https://doi.org/10.48550/ARXIV.1409.8152}
\showDOI{\tempurl}


\bibitem[\protect\citeauthoryear{Mejova, Zhang, Diakopoulos, and
  Castillo}{Mejova et~al\mbox{.}}{2014b}]%
        {Mejova2014ControversyAS}
\bibfield{author}{\bibinfo{person}{Yelena Mejova}, \bibinfo{person}{Amy~X.
  Zhang}, \bibinfo{person}{Nicholas~A. Diakopoulos}, {and}
  \bibinfo{person}{Carlos Castillo}.} \bibinfo{year}{2014}\natexlab{b}.
\newblock \showarticletitle{Controversy and Sentiment in Online News}.
\newblock \bibinfo{journal}{\emph{ArXiv}}  \bibinfo{volume}{abs/1409.8152}
  (\bibinfo{year}{2014}).
\newblock


\bibitem[\protect\citeauthoryear{Morales, Borondo, Losada, and Benito}{Morales
  et~al\mbox{.}}{2015}]%
        {Morales2015MeasuringPP}
\bibfield{author}{\bibinfo{person}{Alfredo~J. Morales}, \bibinfo{person}{Javier
  Borondo}, \bibinfo{person}{Juan~Carlos Losada}, {and}
  \bibinfo{person}{Rosa~M. Benito}.} \bibinfo{year}{2015}\natexlab{}.
\newblock \showarticletitle{Measuring Political Polarization: Twitter shows the
  two sides of Venezuela}.
\newblock \bibinfo{journal}{\emph{Chaos}}  \bibinfo{volume}{25 3}
  (\bibinfo{year}{2015}), \bibinfo{pages}{033114}.
\newblock


\bibitem[\protect\citeauthoryear{Nelson and Webster}{Nelson and
  Webster}{2017}]%
        {doi:10.1177/2056305117729314}
\bibfield{author}{\bibinfo{person}{Jacob~L. Nelson} {and}
  \bibinfo{person}{James~G. Webster}.} \bibinfo{year}{2017}\natexlab{}.
\newblock \showarticletitle{The Myth of Partisan Selective Exposure: A Portrait
  of the Online Political News Audience}.
\newblock \bibinfo{journal}{\emph{Social Media + Society}} \bibinfo{volume}{3},
  \bibinfo{number}{3} (\bibinfo{year}{2017}),
  \bibinfo{pages}{2056305117729314}.
\newblock
\urldef\tempurl%
\url{https://doi.org/10.1177/2056305117729314}
\showDOI{\tempurl}
\showeprint{https://doi.org/10.1177/2056305117729314}


\bibitem[\protect\citeauthoryear{Nyhan and Reifler}{Nyhan and Reifler}{2010}]%
        {Nyhan2010}
\bibfield{author}{\bibinfo{person}{Brendan Nyhan} {and} \bibinfo{person}{Jason
  Reifler}.} \bibinfo{year}{2010}\natexlab{}.
\newblock \showarticletitle{When Corrections Fail: The Persistence of Political
  Misperceptions}.
\newblock \bibinfo{journal}{\emph{Political Behavior}} \bibinfo{volume}{32},
  \bibinfo{number}{2} (\bibinfo{date}{01 June} \bibinfo{year}{2010}),
  \bibinfo{pages}{303--330}.
\newblock
\showISSN{1573-6687}
\urldef\tempurl%
\url{https://doi.org/10.1007/s11109-010-9112-2}
\showDOI{\tempurl}


\bibitem[\protect\citeauthoryear{Pariser}{Pariser}{2011}]%
        {10.5555/2029079}
\bibfield{author}{\bibinfo{person}{Eli Pariser}.}
  \bibinfo{year}{2011}\natexlab{}.
\newblock \bibinfo{booktitle}{\emph{The Filter Bubble: What the Internet Is
  Hiding from You}}.
\newblock \bibinfo{publisher}{Penguin Group , The}.
\newblock
\showISBNx{1594203008}


\bibitem[\protect\citeauthoryear{Preo{\c{t}}iuc-Pietro, Liu, Hopkins, and
  Ungar}{Preo{\c{t}}iuc-Pietro et~al\mbox{.}}{2017}]%
        {preotiuc-pietro-etal-2017-beyond}
\bibfield{author}{\bibinfo{person}{Daniel Preo{\c{t}}iuc-Pietro},
  \bibinfo{person}{Ye Liu}, \bibinfo{person}{Daniel Hopkins}, {and}
  \bibinfo{person}{Lyle Ungar}.} \bibinfo{year}{2017}\natexlab{}.
\newblock \showarticletitle{Beyond Binary Labels: Political Ideology Prediction
  of {T}witter Users}. In \bibinfo{booktitle}{\emph{Proceedings of the 55th
  Annual Meeting of the Association for Computational Linguistics (Volume 1:
  Long Papers)}}. \bibinfo{publisher}{Association for Computational
  Linguistics}, \bibinfo{address}{Vancouver, Canada},
  \bibinfo{pages}{729--740}.
\newblock
\urldef\tempurl%
\url{https://doi.org/10.18653/v1/P17-1068}
\showDOI{\tempurl}


\bibitem[\protect\citeauthoryear{Rizoiu, Graham, Zhang, Zhang, Ackland, and
  Xie}{Rizoiu et~al\mbox{.}}{2018}]%
        {Rizoiu_Graham_Zhang_Zhang_Ackland_Xie_2018}
\bibfield{author}{\bibinfo{person}{Marian-Andrei Rizoiu},
  \bibinfo{person}{Timothy Graham}, \bibinfo{person}{Rui Zhang},
  \bibinfo{person}{Yifei Zhang}, \bibinfo{person}{Robert Ackland}, {and}
  \bibinfo{person}{Lexing Xie}.} \bibinfo{year}{2018}\natexlab{}.
\newblock \showarticletitle{DebateNight: The Role and Influence of Socialbots
  on Twitter During the 1st 2016 U.S. Presidential Debate}.
\newblock \bibinfo{journal}{\emph{Proceedings of the International AAAI
  Conference on Web and Social Media}} \bibinfo{volume}{12},
  \bibinfo{number}{1} (\bibinfo{date}{Jun.} \bibinfo{year}{2018}).
\newblock
\urldef\tempurl%
\url{https://ojs.aaai.org/index.php/ICWSM/article/view/15029}
\showURL{%
\tempurl}


\bibitem[\protect\citeauthoryear{Rogowski and Sutherland}{Rogowski and
  Sutherland}{2016}]%
        {polarizationref1}
\bibfield{author}{\bibinfo{person}{Jon Rogowski} {and}
  \bibinfo{person}{Joseph~L. Sutherland}.} \bibinfo{year}{2016}\natexlab{}.
\newblock \showarticletitle{How Ideology Fuels Affective Polarization}.
\newblock \bibinfo{journal}{\emph{Political Behavior}}  \bibinfo{volume}{38}
  (\bibinfo{date}{06} \bibinfo{year}{2016}).
\newblock
\urldef\tempurl%
\url{https://doi.org/10.1007/s11109-015-9323-7}
\showDOI{\tempurl}


\bibitem[\protect\citeauthoryear{Salloum, Chen, and Kivel\"{a}}{Salloum
  et~al\mbox{.}}{2022}]%
        {10.1145/3512962}
\bibfield{author}{\bibinfo{person}{Ali Salloum}, \bibinfo{person}{Ted Hsuan~Yun
  Chen}, {and} \bibinfo{person}{Mikko Kivel\"{a}}.}
  \bibinfo{year}{2022}\natexlab{}.
\newblock \showarticletitle{Separating Polarization from Noise: Comparison and
  Normalization of Structural Polarization Measures}.
\newblock \bibinfo{journal}{\emph{Proc. ACM Hum.-Comput. Interact.}}
  \bibinfo{volume}{6}, \bibinfo{number}{CSCW1}, Article
  \bibinfo{articleno}{115} (\bibinfo{date}{apr} \bibinfo{year}{2022}),
  \bibinfo{numpages}{33}~pages.
\newblock
\urldef\tempurl%
\url{https://doi.org/10.1145/3512962}
\showDOI{\tempurl}


\bibitem[\protect\citeauthoryear{Tang, Meng, Nguyen, Mei, and Zhang}{Tang
  et~al\mbox{.}}{2014}]%
        {pmlr-v32-tang14}
\bibfield{author}{\bibinfo{person}{Jian Tang}, \bibinfo{person}{Zhaoshi Meng},
  \bibinfo{person}{Xuanlong Nguyen}, \bibinfo{person}{Qiaozhu Mei}, {and}
  \bibinfo{person}{Ming Zhang}.} \bibinfo{year}{2014}\natexlab{}.
\newblock \showarticletitle{Understanding the Limiting Factors of Topic
  Modeling via Posterior Contraction Analysis}. In
  \bibinfo{booktitle}{\emph{Proceedings of the 31st International Conference on
  Machine Learning}} \emph{(\bibinfo{series}{Proceedings of Machine Learning
  Research}, Vol.~\bibinfo{volume}{32})},
  \bibfield{editor}{\bibinfo{person}{Eric~P. Xing} {and} \bibinfo{person}{Tony
  Jebara}} (Eds.). \bibinfo{publisher}{PMLR}, \bibinfo{address}{Bejing, China},
  \bibinfo{pages}{190--198}.
\newblock
\urldef\tempurl%
\url{https://proceedings.mlr.press/v32/tang14.html}
\showURL{%
\tempurl}


\bibitem[\protect\citeauthoryear{Uchida, Toriumi, and Sakaki}{Uchida
  et~al\mbox{.}}{2019}]%
        {uchida}
\bibfield{author}{\bibinfo{person}{Kazuki Uchida}, \bibinfo{person}{Fujio
  Toriumi}, {and} \bibinfo{person}{Takeshi Sakaki}.}
  \bibinfo{year}{2019}\natexlab{}.
\newblock \showarticletitle{Comparative evaluation of two approaches for
  retweet clustering: A text-based method and graph-based method}.
\newblock \bibinfo{journal}{\emph{Web Intelligence}}  \bibinfo{volume}{17}
  (\bibinfo{date}{11} \bibinfo{year}{2019}), \bibinfo{pages}{1--14}.
\newblock
\urldef\tempurl%
\url{https://doi.org/10.3233/WEB-190418}
\showDOI{\tempurl}


\bibitem[\protect\citeauthoryear{Wong, Tan, Sen, and Chiang}{Wong
  et~al\mbox{.}}{2016}]%
        {7454756}
\bibfield{author}{\bibinfo{person}{Felix Ming~Fai Wong},
  \bibinfo{person}{Chee~Wei Tan}, \bibinfo{person}{Soumya Sen}, {and}
  \bibinfo{person}{Mung Chiang}.} \bibinfo{year}{2016}\natexlab{}.
\newblock \showarticletitle{Quantifying Political Leaning from Tweets,
  Retweets, and Retweeters}.
\newblock \bibinfo{journal}{\emph{IEEE Transactions on Knowledge and Data
  Engineering}} \bibinfo{volume}{28}, \bibinfo{number}{8}
  (\bibinfo{year}{2016}), \bibinfo{pages}{2158--2172}.
\newblock
\urldef\tempurl%
\url{https://doi.org/10.1109/TKDE.2016.2553667}
\showDOI{\tempurl}


\bibitem[\protect\citeauthoryear{Xiao, Song, Xu, Ren, and Sun}{Xiao
  et~al\mbox{.}}{2020}]%
        {10.1145/3394486.3403275}
\bibfield{author}{\bibinfo{person}{Zhiping Xiao}, \bibinfo{person}{Weiping
  Song}, \bibinfo{person}{Haoyan Xu}, \bibinfo{person}{Zhicheng Ren}, {and}
  \bibinfo{person}{Yizhou Sun}.} \bibinfo{year}{2020}\natexlab{}.
\newblock \showarticletitle{TIMME: Twitter Ideology-Detection via Multi-Task
  Multi-Relational Embedding}. In \bibinfo{booktitle}{\emph{Proceedings of the
  26th ACM SIGKDD International Conference on Knowledge Discovery \& Data
  Mining}} (Virtual Event, CA, USA) \emph{(\bibinfo{series}{KDD '20})}.
  \bibinfo{publisher}{Association for Computing Machinery},
  \bibinfo{address}{New York, NY, USA}, \bibinfo{pages}{2258–2268}.
\newblock
\showISBNx{9781450379984}
\urldef\tempurl%
\url{https://doi.org/10.1145/3394486.3403275}
\showDOI{\tempurl}


\bibitem[\protect\citeauthoryear{Yang, Wen, Lin, and Deng}{Yang
  et~al\mbox{.}}{2017}]%
        {8181508}
\bibfield{author}{\bibinfo{person}{Muheng Yang}, \bibinfo{person}{Xidao Wen},
  \bibinfo{person}{Yu-Ru Lin}, {and} \bibinfo{person}{Lingjia Deng}.}
  \bibinfo{year}{2017}\natexlab{}.
\newblock \showarticletitle{Quantifying Content Polarization on Twitter}. In
  \bibinfo{booktitle}{\emph{2017 IEEE 3rd International Conference on
  Collaboration and Internet Computing (CIC)}}. \bibinfo{pages}{299--308}.
\newblock
\urldef\tempurl%
\url{https://doi.org/10.1109/CIC.2017.00047}
\showDOI{\tempurl}


\end{thebibliography}

\appendix

\end{document}